\documentclass[prb, aps, reprint]{revtex4-1}
\pdfoutput=1

\usepackage{lmodern}
\usepackage[english]{babel} 
\usepackage{graphicx}
\usepackage{color}  
\usepackage[draft]{changes} 
\usepackage{subcaption}
\usepackage[space]{grffile}
\definecolor{navyblue}{rgb}{0.0, 0.0, 0.5} 
\usepackage{amsmath, amssymb, amsfonts,bm}
\usepackage{latexsym}   
\usepackage{bbm} 
\usepackage{mathtools}  
\usepackage{placeins}   
\usepackage[protrusion=true, expansion=true]{microtype}
\usepackage[colorlinks=true, breaklinks=false, linkcolor=navyblue, urlcolor=navyblue, citecolor=navyblue]{hyperref}

\newcommand{\ket}[1]{\mathinner{|{#1}\rangle}}

\newcommand{\braket}[2]{\langle #1 | #2 \rangle}

\newcommand{\expe}[1]{\left\langle #1 \right\rangle}

\begin{document}
\title{Parametric resonance of Josephson plasma waves:
A theory for optically amplified interlayer superconductivity in $\rm{YBa_2Cu_3O_{6+x}}$}

\author{Marios H. Michael$^1$}\email{marios\_michael@g.harvard.edu}
\author{Alex von Hoegen$^2$}
\author{Michael Fechner$^2$}
\author{Michael F\"{o}rst$^2$}
\author{Andrea Cavalleri$^{2,3}$}
\author{Eugene Demler$^1$}
\affiliation{$^1$Department of Physics, Harvard University, Cambridge, Massachusetts, USA}
\affiliation{$^2$Max Planck Institute for the Structure and Dynamics of Matter, Hamburg, Germany}
\affiliation{$^3$Department of Physics, University of Oxford, UK}

\date{\today}
\selectlanguage{english}
\begin{abstract}
Non-linear interactions between collective modes play a definitive role in far out of equilibrium dynamics of strongly correlated electron systems. Understanding and utilizing these interactions is crucial to photo-control of quantum many-body states. One of the most surprising examples of strong mode coupling is the interaction between apical oxygen phonons and Josephson plasmons in bilayer $\rm{YBa_2Cu_3O_{6+x}}$ superconductors. Experiments by Hu et al.\cite{Hu14} and  Kaiser et al.\cite{Kaiser14} showed that below $T_c$, photo-excitation of phonons leads to enhancement and frequency shifts of Josephson plasmon edges, while above $T_c$, photo-excited phonons induce plasmon edges even when there are no discernible features in the equilibrium  reflectivity spectrum.  Recent experiments by Van Hoegen et al. \cite{vanHoegen19} also observed parametric generation of Josephson plasmons from photo-excited phonons both below $T_c$ and in the pseudogap phase. In this paper we present a theoretical model of phonon-plasmon three wave interaction arising from coupling between the oxygen motion and the in-plane superfluid stiffness. Analysis of the parametric instability of plasmons based on this model gives frequencies of the most unstable plasmons that are in agreement with experimental observations. We also discuss how strong parametric excitation of Josephson plasmons can explain pump induced changes in the TeraHertz reflectivity of $\rm{YBa_2Cu_3O_{6+x}}$ in the superconducting state, including frequency shifts and sharpening of Josephson plasmon edges, as well as appearance of a new peak around 2THz. An interesting feature of this model is that overdamped Josephson plasmons do not give any discernible features in reflectivity in equilibrium, but can develop plasmon edges when parametrically excited. We suggest that this mechanism explains photo-induced plasmon edges in the pseudogap phase of $\rm{YBa_2Cu_3O_{6+x}}$.
\end{abstract}

\maketitle

% {\bf I changed symmetry definition from "parity inversion" to reflection in the plane symmetry. Check if you agree and whether this introduced any inconsistencies. Put $()$ when referencing equations.
% \newline
% Should we introduce abbreviation AO for "apical oxygen" and use it everywhere?
% \newline
% When parametric resonance involves creation of plasmons in different plasmon bands, I am not sure that the argument of "half the driving frequency is correct". Symmetry considerations suggest that when driving comes from the upper JP, it generates plasmons in different bands (is this correct?). Hence our argument for renormalization of the upper JP may be suspect. 
% }
\section{Introduction}

\subsection{Motivation}
\begin{figure}[t]
\center
\includegraphics[trim={1.5cm 4cm 2cm 3cm},clip, scale=0.7]{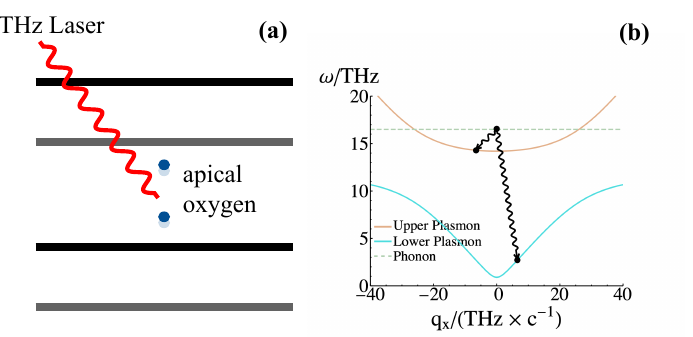}
\caption{a) Schematic representation of the pumping scheme. A laser pulse at frequency $\omega = 16.5$ THz coherently excites the resonant phonon corresponding to oscillations of the apical oxygen sitting in between the two layers of the bilayer structure. b) The orange and blue lines show dispersions of the upper and lower Josephson polaritons as a function of in-plane momentum (see section \ref{sec:model} for details). The phonon mode at $q_{ph.}= 0$ can be resonantly scattered into a pair of JPs with opposite momenta. The resonant scattering condition is $\omega_{\rm ph} = \omega_1(\vec{q}) + \omega_2 (-\vec{q})$. The figure shows the resonant process at $q_z =0$ only. There is a full 2D surface of resonant processes with the parametric instability for different pairs determined by the corresponding matrix elements (see Fig.~\ref{fig:intersec} and Section \ref{sec:instability} for details). The dominant instability occurs for  $q_z$ near the Brillouin zone boundary. }
\label{fig:1upper1lower}
\end{figure}

Photo-induced superconductivity has been one of the most surprising discoveries of pump and probe experiments in solids. By now it has been observed in three different classes of materials: high $T_c$ cuprates \cite{Hu14, Kaiser14, vanHoegen19, Fausti11,Nicoletti2014,Zhang2018,Cremin2019}, the buckyball superconductor $K_3C_{60}$\cite{budden20}, and organic superconductors such as $(BEDT-TTF)_2 Cu[N(CN)_2]Br$ \cite{buzzi20}. In many of these materials, superconductivity has been induced by tuning the pump pulse to be resonant with one of the IR active phonon modes; interaction between the strongly excited phonon modes and the many-body electron system is at the heart of this phenomenon. The primary evidence for transient superconductivity comes from reflectivity measurements, which allow to study optical conductivity $\sigma(\omega)$. Suppression of the real part of the conductivity at low frequencies indicates opening of the quasiparticle gap, and appearance of the $1/\omega$ tail in the imaginary part suggests the appearance of finite frequency superfluid stiffness. In the case of high $T_c$ superconductors, and especially $\rm{YBa_2Cu_3O_{6+x}}$, another important indication of the photo-induced superconducting state has been the appearance of special features in the reflection spectra, called the Josephson plasmon (JP) edge \cite{Hu14}\cite{Kaiser14} (see section \ref{sec:statrefl} for a detailed discussion). In equilibrium, JPs are understood to be a consequence of coherent Cooper pair tunneling between the layers, and thus provide a hallmark signature of the superconducting state. 
Experiments by Hu et al.\cite{Hu14} and  Kaiser et al.\cite{Kaiser14} showed that following photo-excitation of apical oxygen (AO) phonons in $\rm{YBa_2Cu_3O_{6+x}}$(Fig.~\ref{fig:1upper1lower}a), JP edges get strongly enhanced: they sharpen below $T_c$ and may appear out of the featureless optical conductivity above $T_c$. 
Moreover, consequences of pumping include the appearance of an additional photo-induced edge below $T_c$ and a shift in frequency of JP edges relative to their equilibrium values both below and above $T_c$. The appearance of this additional blue shifted lower plasmon edge has been interpreted as indication of the photo-induced enhancement of the interlayer Josephson coupling. Earlier theoretical work aimed at understanding photo-induced superconductivity\cite{Dasari18,Babadi17,Yuta17,Okamoto17,Denny15,Dante19} has not been able to explain the dramatic enhancement of JP edges following excitation of the phonons. Another striking phenomenon that has been recently observed in $\rm{YBa_2Cu_3O_{6+x}}$ is the coherent excitation of Josephson plasma wave oscillations by the photo-excited phonons\cite{vanHoegen19}. Intriguingly, such resonant excitation of plasmons has been observed not only below $T_c$, but also in the pseudogap phase. The goal of this paper is to provide a unified physical model for the phonon-plasmon interaction in $\rm{YBa_2Cu_3O_{6+x}}$ that explains all these different experimental signatures.

\subsection{Nontechnical summary of the paper}

The starting point of our theoretical study is the experimental 
finding by Van Hoegen et al. \cite{vanHoegen19} that illuminating the superconducting $\rm{YBa_2Cu_3O_{6+x}}$ with 
light resonant with AO phonons leads to the appearance of coherent Josephson plasmon excitations. 
These experiments used
Second Harmonic Generation (SHG) measurements at optical frequencies
to probe time dependent breaking of inversion symmetry, which can arise from both
phonon displacements and electric currents. 
From the perspective of THz dynamics of collective modes, measurements at the 
optical frequency are essentially instantaneous. Thus the SHG signal
allows measurement of the time-dependence of inversion symmetry breaking, which 
can then be Fourier transformed to determine frequencies of the 
excited collective modes. Experiments revealed strong signals at
frequencies corresponding to lower and upper Josephson plasmons
not only below Tc, but also in the pseudogap phase. This was interpreted as
a result of
resonant parametric scattering between the photoexcited AO phonon and
two Josephson plasmons in a three-wave mixing process shown in Fig.~\ref{fig:1upper1lower}. 
Further support for this model came from the measured exponential dependence
of the induced JP currents on the pump amplitude. In this paper we
present a detailed model of
phonon-plasmon resonant parametric scattering in $\rm{YBa_2Cu_3O_{6+x}}$, in which phonon-plasmon coupling 
comes from the modulation of in-plane superfluid density in Cu-O layers 
by apical oxygen phonons. 
We show that while the parametric resonance condition is satisfied for a range of upper/lower Josephson plasmon
pairs, the most unstable pair has frequencies of 2.2 THz and 14.3 THz. These frequencies are in good agreement
with the experimentally measured values \cite{vanHoegen19}. 

The second goal of this paper is to show that non-linear couplings between
collective modes, including both phonon-plasmon and plasmon-plasmon interactions, provide a natural explanation for previously observed changes in the terahertz reflectivity.
To understand these phenomena we need to broaden the discussion of
parametric resonant scattering processes to include second order processes, in which phonon-induced 2.2 THz and 14.3 THz plasmons
themselves become a source of  parametric driving.
While there are many possibilities for higher order
parametric instabilities, we can identify the ones that are most important for understanding the renormalization
of optical reflectivity from considerations of energy and momentum conservation. 
For example, we find that reflectivity close to the lower plasmon
edge is renormalized either by a four wave mixing of
two neighbouring phonons and two lower Josephson plasmons,
as shown in Fig. 3a, or a four wave mixing between
four lower plasmons as shown in Fig.~\ref{fig:4waveplasmons} (discussed in
Sec. V). Both mechanisms give rise to similar features in
reflectivity, shown in Fig.~\ref{fig:4wave}b.

The main focus of our paper is understanding pump-induced changes in the
terahertz reflectivity of $\rm{YBa_2Cu_3O_{6+x}}$ in the superconducting state. We demonstrate that parametric 
driving can explain renormalization of JP energies, sharpening of the edges, and 
appearance of the new edge around 2 THz. We also
provide a qualitative discussion of the pseudogap regime, which we envision as having short-ranged superconducting correlations \cite{Uemura91,Emery95,CHEN05,LOKTEV01,Tesanovic02,Wang01,Podolsky07,Li10,Keimer15,Zhou19}.
Earlier analysis of Josephson plasmons in the vortex liquid state of high Tc cuprates 
by Bulaevskii and Koshelev \cite{Koshelev2014} has already established that
finite correlations between neighboring layers are sufficient for observing plasmons 
as collective polaritonic modes. The main consequence of the absence of
long-range superconducting order is enhanced broadening of the plasmon resonance.
Furthermore, in the pseudogap regime, one expects higher damping of plasmons. This damping comes from 
an increase in the real part of optical conductivity
at the plasmon frequency; in the superconducting state below the frequency of the quasiparticle continuum, the
real part of conductivity of d-wave superconductors is strongly suppressed \cite{Won94}.
In this spirit we treat Josephson plasmons in the pseudogap regime as overdamped excitations
that undergo parametric driving following photoexcitation. We demonstrate that resonant parametric driving 
can explain the emergence of the lower JP edge out of the featureless continuum in reflectivity data; This is the most striking feature of the "light induced superconductivity" in the pseudogap regime of $\rm{YBa_2Cu_3O_{6+x}}$: 

\begin{figure}[t]
    \center
     \includegraphics[trim={2.5cm 8cm 1cm 0cm},clip, scale=0.65]{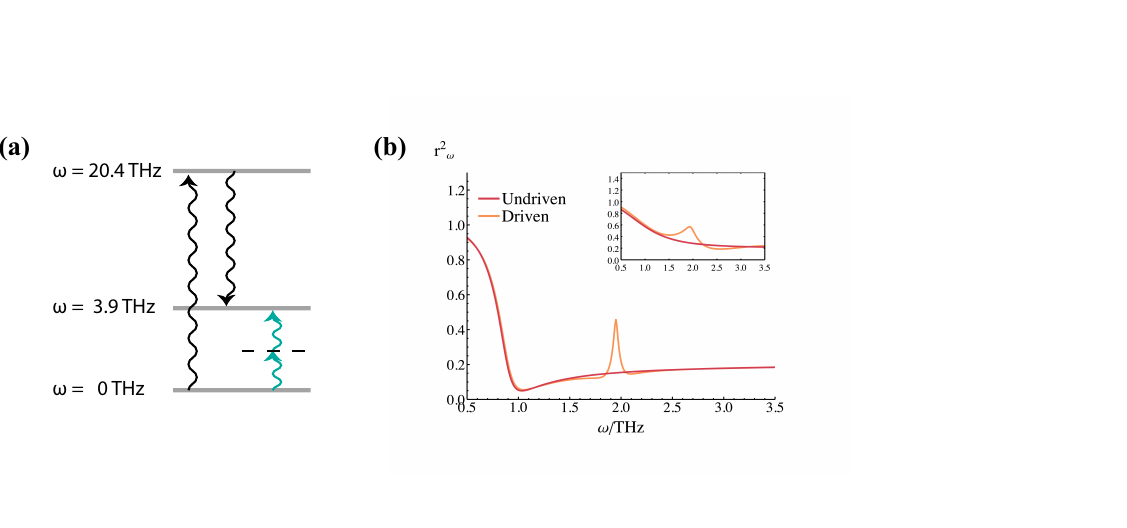}
    \caption{a) Schematic representation of the four wave mixing process renormalizing the reflection coefficient. In this four wave mixing process the sum of a pair of plasmons (green) adds up to the subtraction of neighbouring phonon frequencies (black). b) Reflection coefficient with and without driving. Inset: Emergence of a photo-induced edge from a featureless background in the case of strongly dissipative plasmons.}
    \label{fig:4wave}
\end{figure}

\subsection{Organization of the paper}

\textit{Josephson plasmons and terahertz reflectivity in equilibrium}. We begin Section~\ref{sec:model} by presenting a microscopic theory of JPs in bilayer superconductors, such as $\rm{YBa_2Cu_3O_{6+x}}$. Our analysis is based on combining hydrodynamic equations of the superconducting order parameter with Maxwell equations for electric and magnetic fields. This approach is equivalent to the linearized version of the three dimensional sine-Gordon type
equation of  electrodynamics of layered superconductors obtained in Refs. \cite{Kleiner94,Bulaevskii94,Bulaevskii96,Koshelev2000}. In contrast to these papers we retain electric and magnetic fields as explicit degrees of freedom, in order to facilitate identification of eigenmodes, analysis of resonant scattering processes, and computation of the frequency dependent reflectivity.   We conclude the discussion of the equilibrium properties of plasmons by presenting a simple analytical model for calculating reflectivity in a bi-layer superconductor and show that results of this analysis are in good agreement with experiments. 
In Section ~\ref{sec:phonon} we introduce AO phonon modes and review symmetry constraints on their non-linear coupling to Josephson plasmons. We suggest that the microscopic origin of phonon-plasmon interaction is modulation of the in-plane superfluid stiffness by the apical oxygen displacements, coming from modifications either in the density or the effective mass of the charge carriers. It is useful to note that phonon-plasmon coupling can not be captured using the approach of Refs. \cite{Marel01,Koyama96,Koyama2002,Okamoto16,Okamoto17}, which treats a layered superconductor as a chain of Josephson junctions\cite{Marel01,Bulaevskii94,Koyama96,Koyama2002,Okamoto16,Okamoto17,Laplace2016} stacked along the $c$-axis, thus neglecting the in-plane superfluid currents. Due to symmetry constraints
three-wave phonon-plasmon coupling only takes place for finite in-plane momentum of plasmons.

\textit{Parametric instabilities and terahertz reflectivity in a driven state}.
Section~\ref{sec:instability} presents  analysis of the parametric instability in which one photo-excited phonon generates a pair of Josephson polaritons that have opposite momenta and energies that add-up to the phonon frequency (see Fig.~\ref{fig:1upper1lower}). The parametric resonance condition does not select  a unique pair of plasmons, but is satisfied for JPs on a 2D surface in momentum space. We calculate the growth rate for all resonant pairs and find that the most unstable modes have frequencies 2.2 and 14.3 THz, in agreement with the Second Harmonic Generation experiments\cite{vanHoegen19}. In section ~\ref{sec:refl} we extend analysis of the terahertz reflectivity
of bilayer superconductors to driven systems using the perspective of Floquet medium \cite{Sho19,Buzzi19}. In this discussion sources of the drive can be either the original photo-excited phonons, or the unstable plasmon modes identified in section \ref{sec:instability} (see Ref. \cite{Rajasekaran2016} for a discussion on considering the incoming electric field from the pump field directly as the source of the drive). For each frequency window considered, dominant contributions come from processes that are closest to parametric resonance and consistent with symmetry selection rules. In the frequency region close to the lower JP edge, the above considerations suggests two mechanisms that can lead to renormalization of the reflection coefficient: i) a four wave mixing process that involves two different AO phonons and two lower JPs, ii) a four wave mixing process between the lower JPs, in which two plasmons have been generated  by the phonon decay. Both modulation mechanisms are expected to result in a similar change of the reflection coefficient, with the strongest feature being a new peak at the frequency around 2 THz. Reflection at frequencies close to the upper JP edge are modified by the four plasmon interaction, where two plasmons have been generated in the decay of the photoexcited phonon. 
On a technical level, theoretical tools developed in this section represent a generalization of previous analysis of reflectivity from a Floquet medium \cite{Sho19,Buzzi19} to the case of multiple plasmon-polariton bands corresponding to different transmission/reflection channels for light within the material. We decompose  electromagnetic fields and matter excitations into eigenstates corresponding to different plasmon-polariton bands and project light reflection problem to the relevant eigenstate subspace for the frequency window of interest. Within that subspace we include only resonant time-dependent processes, equivalent to a first order Floquet degenerate perturbation scheme\cite{Eckardt15}. In practice this means keeping only {\it signal} and {\it idler} frequency components for the different transmission channels that we consider\cite{Schackert13}. Our analysis does not include depletion of the driving modes and therefore only captures short-time dynamics of the pump and probe experiments.  In section ~\ref{sec:refl} we also demonstrate that in systems with overdoped JPs, which do not show plasma edges in equilibrium, parametric driving can lead to the appearance of the edge. We suggest that this can explain light induced superconductivity in the pseudogap phase of $\rm{YBa_2Cu_3O_{6+x}}$.
In section~\ref{sec:concl}, we summarize our results and review  open problems.

\section{Theoretical analysis of Josephson plasmons in bilayer superconductors}
\label{sec:model}

\begin{figure}[t]
    
     \includegraphics[trim={6cm 5cm 0cm 5cm},clip, scale=0.9]{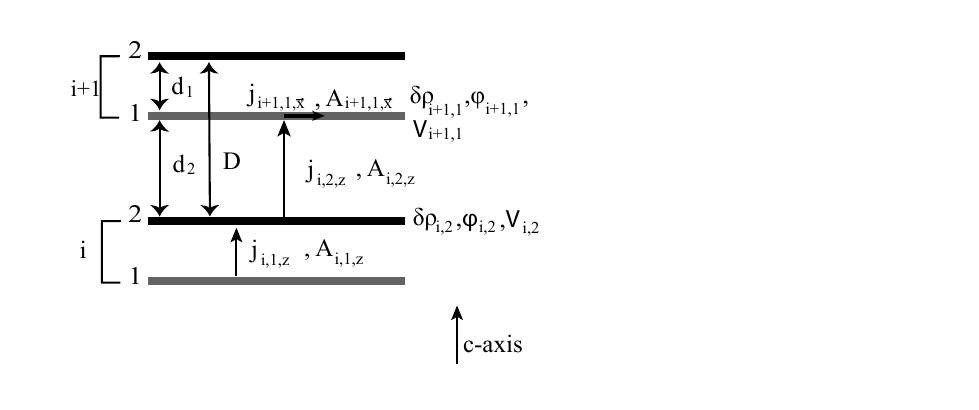}
    \caption{Schematic view of the $\rm{YBa_2Cu_3O_{6+x}}$ material. $\rm{YBa_2Cu_3O_{6+x}}$ consists of a bilayer array of superconducting $CuO_2$-layers. Each layer is described by the density and phase fluctuations of the superconductor. Electric fields and currents along the c-axis live in the space between the layers while, in-plane electric fields and currents live inside the layer.}
    \label{fig:bilayer}
\end{figure}

We analyze dynamics of a bilayer system such as $\rm{YBa_2Cu_3O_{6+x}}$ by developing a microscopic description based on phase and density fluctuations in each superconducting layer $\{\delta \rho_{i,\lambda}, \phi_{i,\lambda} \}$, coupled to the 4-component vector potential $(\,V_{\lambda,i}(\vec{x}), {A}_{\lambda,i,z}(\vec{x})\, \vec{A}_{\lambda,i,\vec{x}}(\vec{x})\,)$ through Maxwell's equations. Here  $i$ corresponds to the index of the unit cell along the c-axis, $\lambda=1,2$ labels the number of the layer inside the unit cell, and $\vec{x}$ is the in-plane coordinate, which we
will omit in the equations below for brevity. While the in-plane components of the vector potential $\vec{A}_{\lambda,i,\vec{x}}(\vec{x})$ are defined within the corresponding layers, ${A}_{\lambda,i,z}(\vec{x})$ is defined to be on the links between layers starting on layer $\{\lambda,i\}$ as shown in Fig.~\ref{fig:bilayer}. Our analysis focuses on density and current fluctuations, since we expect super flow currents to dominate at frequencies below $2 \Delta$. In section~\ref{sec:statrefl}, we add dissipation phenomenologically to include effects of quasiparticle excitations. Equations of motion are given by hydrodynamic equations discretized along the $c$-axis. Our treatment expands on ideas developed in previous works treating $\rm{YBa_2Cu_3O_{6+x}}$ as a 1D chain of Josephson junctions\cite{Marel01,Bulaevskii94,Koyama96,Koyama2002,Okamoto16,Okamoto17} to include in-plane currents in a consistent way. Dynamics of density fluctuations is given by the continuity equation,
\begin{equation}
\partial_t \delta \rho_{\lambda,i}+ \partial_x j_{\lambda,i,x}+ \Delta_z j_{\lambda,i,z}  = 0.
\label{eq:continuity}
\end{equation}
In linearized hydrodynamics the supercurrents are given by \cite{currentdeff20}
\begin{subequations}
	\begin{align}
	&j_{\lambda,i,x} = \Lambda_s \left( \partial_x \phi_{\lambda,i} - e^* A_{\lambda,i,x} \right), \\
	&j_{\lambda,i,z} = j_{c\lambda} d_\lambda^2 \left( \Delta_z \phi_{\lambda,i} - e^*A_{\lambda,i,z} \right).
	\end{align}
	\label{eq:curr}
\end{subequations}
Here $\vec{x}$ denotes the in-plane $x,y$ components and $z$-axis denotes the  c-axis of the crystal, coupling to the vector potential is given by the Cooper pair charge, $e^* = 2e$ and we work in units where $\hbar  = 1$ for the rest of the paper.
The in-plane components of the superfluid current are defined within individual layers and have continuous gradients. The $z$-component of the current is defined as the Josephson current between adjacent layers and has a lattice gradient which corresponds to the phase difference between adjacent layers,
\begin{equation}
\begin{split}
\Delta_z \phi_{\lambda,i}=  \left\{ \begin{array} {c} (\phi_{2,i} -  \phi_{1,i})/d_1, \mbox{ for } \lambda =1,\\
(\phi_{1,i+1} -  \phi_{2,i})/d_2, \mbox{ for } \lambda =2
\end{array} \right.
\end{split}    
\end{equation}
In the continuity equation (\ref{eq:continuity}), the lattice gradient of $j_{\lambda,i,z}$ (defined on links) corresponds to the difference of currents flowing in and out of the $\{\lambda,i\}$-layer and is given by\cite{currentdeff20}
\begin{equation}
\begin{split}
\Delta_z j_{\lambda,i,z} = \left\{ \begin{array} {c} \frac{j_{1,i,z}}{d_1} -  \frac{j_{2,i-1,z}}{d_2}, \mbox{ for } \lambda =1,
\\  \frac{j_{2,i,z}}{d_2} -  \frac{j_{1,i,z}}{d_1}, \mbox{ for } \lambda =2
\end{array} \right.
\end{split}  
\label{J_zgrad_def}  
\end{equation}
Coefficient $\Lambda_s$ is related to the in-plane London penetration length as $\Lambda_s = \frac{   \epsilon c^2 }{\lambda_L^2 (e^*)^2}$, where $\epsilon = \epsilon_r\epsilon$ the permittivity of the material. Physically, it corresponds to the intra-layer superfluid stiffness and is proportional to the condensate density, $\Lambda_{s\lambda} \propto \rho_{\lambda}$. In linear analysis of collective modes we can set $\Lambda _{s\lambda}$ to be equal to their equilibrium values since they multiply superfluid velocities, $\vec{v}_{\lambda,i} = \partial_{\vec{x}}  \phi_{\lambda,i} - e^*A_{\lambda,i,\vec{x}}$, which are already first order
in fluctuations. This is why we omitted the layer index  for $\Lambda_s$ in equation (\ref{eq:curr}).  Coefficients $\{j_{c,\lambda}\}$ correspond to interlayer Josephson tunneling couplings and obey $j_{c,\lambda} \propto \sqrt{\rho_{1} \rho_{2}} $. In linearized hydrodynamics, we take $j_{c,\lambda}$ to be equal to their equilibrium value and neglect corrections due to  $\delta \rho_\lambda$. Phase dynamics are given through the Josephson relation,
\begin{equation}
\partial_t \phi_{\lambda,i} =  - \gamma \delta \rho_{\lambda,i} - e^* V_{\lambda,i},
\label{eq:joseph}
\end{equation}
In Eq~\ref{eq:joseph}, $\gamma$ arises from a finite electron compressibility\cite{Marel01} and can be linked to the Thomas-Fermi length by $\gamma = \frac{\lambda_{TF}^2 (e^*)^2 }{\epsilon}$.

To consistently couple to electromagnetism, one needs to use discrete QED along the $c$-axis and continuous QED in-plane. In the Lorenz gauge, the Maxwell's equations obtain the familiar form:
\begin{subequations}
\begin{align}
\left( \frac{1}{c^2}\partial_t^2 - \partial_x^2 - \Delta_z^2 \right) V_{\lambda,i} &= \frac{(e^*)}{\epsilon} \delta \rho_{\lambda,i},\\
\left( \frac{1}{c^2} \partial_t^2 - \partial_x^2 - \Delta_z^2 \right) A_{\lambda,i,x} &= \frac{1}{c^2} \frac{(e^*)}{\epsilon} j_{\lambda,i,x}, \\
\left( \frac{1}{c^2} \partial_t^2 - \partial_x^2 - \Delta_z^2 \right) A_{\lambda,i,z} &= \frac{1}{c^2} \frac{(e^*)}{\epsilon} j_{\lambda,i,z} 
\end{align}
\label{eq:photon}
\end{subequations}
where along the $c$-axis, lattice gradients are used and the magnetic permeability, $\mu = \mu_r \mu_0$ is included through the speed of light in the material, $c = \frac{1}{\sqrt{\mu \epsilon}}$. The discretized Lorenz gauge condition is given by
\begin{equation}
\frac{1}{c^2}\partial_t V_{\lambda,i} + \partial_x A_{\lambda,i,x} + \Delta_z A_{\lambda,i,z}(x,t) = 0,
\label{eq:gaugecond}
\end{equation}
With the constraint in equation~(\ref{eq:gaugecond}), the above equations of motion (\ref{eq:continuity}-\ref{eq:photon}) can be compactly encoded by the Hamiltonian
\begin{equation}
\begin{split}
H =& \int d^2 x \sum_{i, \lambda} \bigg\{ \frac{\gamma}{2} \delta \rho_{\lambda,i}^2 +\frac{1}{2 \Lambda_s} j_{\lambda,i,x}^2 + \\
& \frac{1}{2 j_{c,\lambda} d^2_\lambda} j^2_{\lambda,i,z} + e^* \delta \rho_{\lambda,i}V_{\lambda,i} + H_{EM} \bigg\},\\
&
\\&
\end{split}
\label{eq:hamil}
\end{equation}
where $H_{EM}$ gives rise to Maxwell's equations of motion. In the Lorenz gauge this Hamiltonian takes the form:
\begin{widetext}
\begin{equation}
\begin{split}
    H_{EM} =& \int d^2 x \sum_{i, \lambda}  \frac{ c^2}{2 \epsilon} P_{V,\lambda,i}^2 + \frac{\epsilon}{2} \left( \left(\partial_x V_{\lambda,i}\right)^2 + \left(\Delta_z V_{\lambda,i}\right)^2\right) \\
    &+  \frac{1}{2 \epsilon} P_{A_x,\lambda,i}^2 + \frac{\epsilon c^2}{2 } \left( \left(\partial_x A_{\lambda,i,x}\right)^2 + \left(\Delta_z A_{\lambda, i, x}\right)^2\right) \\
    &+  \frac{1}{2 \epsilon}\frac{P_{A_z,\lambda, i}^2}{2} + \frac{\epsilon c^2}{2 } \left( \left(\partial_x A_{\lambda,i, z}\right)^2 + \left(\Delta_z A_{\lambda, i, z }\right)^2\right)
\end{split}
\label{eq:photonHamil}
\end{equation}
\end{widetext}
where the fields $\{P_{V,\lambda,i}, P_{A_x,\lambda,i} , P_{A_z,\lambda,i}\}$ have been introduced and correspond to the conjugate momenta of the components of the electrostatic and vector potential. Canonically conjugate pairs, $\rho$ and $\phi $, $V$ and $P_V$, $\vec{A}$ and $P_{\vec{A}}$ obey canonical commutations relations, i.e. $\left[\rho(\vec{x})_i , \phi_j (\vec{x}') \right] = i \delta(\vec{x} - \vec{x}') \delta_{i,j}$. Equations of motion for the operators are recovered through Heisenberg equations of motion, $\partial_t \hat{O} = i \left[ \hat{H},\hat{O}\right]$.

The advantage of using the Lorenz gauge is evident from the fact that different gauge fields decouple from each other (Eq.~(\ref{eq:photon})). The price we pay is that, unlike the Coulomb gauge, the Hamiltonian in equation~(\ref{eq:photonHamil}) has additional unphysical degrees of freedom that can be removed through the gauge condition in equation~(\ref{eq:gaugecond}).

\begin{figure}
\includegraphics[trim={1cm 1cm 0cm 1cm},clip, scale=0.5]{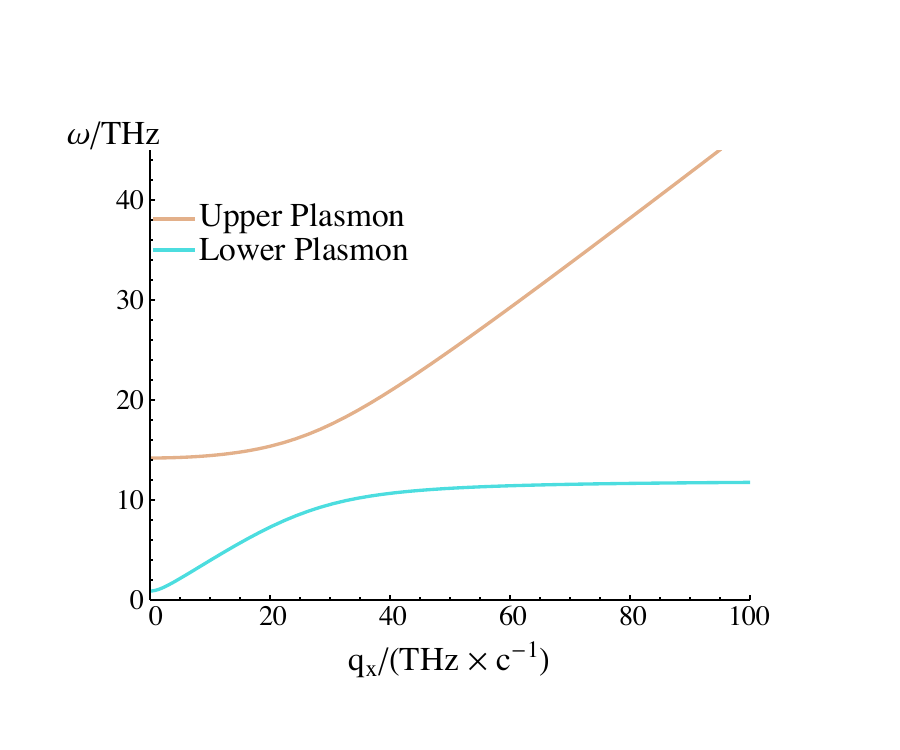}
\caption{Dispersion relation of two lower plasmons. For small wave vectors, $q_x << \frac{\sqrt{\Lambda_s}}{c}$, one can ignore the presence of the in-plane plasmon at equilibrium.}
\label{fig:plasmons}
\end{figure}
\subsection{Quantization in the Lorenz Gauge}
\label{subsec:lorenz}
Quantization in the Lorenz gauge follows a method introduced by Gupta and Bleuler \cite{Gupta50,Bleuler50}. The Hamiltonian is quantized by treating each gauge field as a scalar. This introduces unphysical polarizations, namely a longitudinal and a scalar polarization. These are removed by defining the physical Hillbert space as the set of states that satisfy the Lorenz gauge condition, equation~(\ref{eq:gaugecond}). While the literature typically focuses on QED in free space, we discuss in this subsection how this idea is applied for QED coupled to matter in the particular example presented in this paper.

Starting from the equations of motion (\ref{eq:continuity}-\ref{eq:photon}), dispersion relations of elementary excitations of the system (plasmons) are found by looking for plane wave solutions:
\begin{equation}
\delta \rho_{\lambda,i}(\vec{x},t) = \delta \rho_{\lambda}(q_x,q_z,\omega) e^{ i q_{\vec{x}} \vec{x} + i q_z D i - i \omega t}.
\end{equation}
In the Hamiltonian description of the equations of motion where the momenta of the gauge fields $\{P_V, P_{A_x},P_{A_z} \}$ are included, the equations of motion become first order in the time derivatives and take the form:
\begin{equation}
\begin{split}
    i \partial_t \vec{v} =& M \cdot \vec{v}
\end{split}
\label{eq:eigenEQ}
\end{equation}
where $\underline{v}$ is the vector containing the variables in either layer of $\rm{YBa_2Cu_3O_{6+x}}$, $\vec{v} = \{ \rho_\lambda, V_\lambda, A_{\lambda,x}, A_{\lambda,z}, \phi_\lambda, P_{V,\lambda}, P_{A_x,\lambda}, P_{A_z,\lambda} \} $ and $\lambda = \{1,2\}$. $M$ is a 16x16 matrix coupling the different variables together and contains gradient operators. Substituting the plane wave ansatz this reduces to a 16x16 eigenvalue problem of the type:
\begin{equation}
      \omega \underline{v}(\vec{q},\omega) =M(\vec{q})
    \cdot \underline{v}(\vec{q},\omega).
\end{equation}
The dispersion of elementary excitations are given by solving the corresponding secular equation:
\begin{equation}
    \chi(\omega) = {\rm det} \left( \underline{\underline{M}} - \omega \right) = 0.
\label{eq:secul}
\end{equation}
In equation~(\ref{eq:secul}), $\chi(\omega)$ is the characteristic polynomial of the equations of motion whose solution contains both physical and unphysical energy states. However the gauge condition (\ref{eq:gaugecond}) and continuity equation (\ref{eq:hamil}) guarantee that the unphysical degrees of freedom are decoupled from the matter fields $\{\rho,\phi\}$. This decoupling of matter fields and unphysical degrees of freedom holds for a generic system and not just this example. As a result the characteristic polynomial factorizes into physical and unphysical contributions, $
    \chi(\omega) = \chi_{phys} (\omega) \chi_{unphys}(\omega).
$ Therefore, coupling to matter provides a particularly simple recipe for constructing the physical Hillbert space by ignoring $\chi_{unphys}(\omega)$ and solving only for
\begin{equation}
    \chi_{phys}(\omega) = 0.
\end{equation}
This procedure amounts to explicitly removing the unphysical eigenstates from the theory. Finally, in order to restrict the theory to the physical Hilbert space, variables are expressed as a linear combination of the eigenvectors corresponding to the physical eigenstates only:
\begin{equation}
\begin{split}
    \begin{pmatrix}
    \delta \rho_\lambda(q) \\
    \vdots \\
    \phi_\lambda(q) \\
    \vdots 
    \end{pmatrix}
    = \begin{pmatrix} 
    v^{\delta \rho_\lambda}_{1,q} &&\dots && (v^{\delta \rho_\lambda}_{1,q})^* && \dots \\
   \vdots && && \vdots &&\\
    v^{\phi_\lambda}_{1,q} && && (v^{\phi_\lambda}_{1,q})^* &&  \\
    \vdots && && \vdots &&\\
    \end{pmatrix} \cdot
    \begin{pmatrix} 
    b_1 \\
    \vdots\\
    b_1^* \\
    \vdots
    \end{pmatrix}
\end{split}
\label{eq:eigenDec}
\end{equation}
Quantization of the theory is completed by promoting the coefficients of eigenvectors of the secular equation to creation/annihilation operators with bosonic commutations relations:
\begin{equation}
    \left[ \hat{b}_i, \hat{b}^\dag_{j'} \right] = \delta_{i,j}
\end{equation}
This choice fixes the normalization constant of the eigenvectors as shown and completes our construction as outlined in Appendix~\ref{app:eigen}. Details of the mathematical structure of the diagonalization scheme can be found in Ref.~\cite{xiao09theory} where one can prove that for a real spectrum, the eigenvectors form an orthonormal basis of the physical system. Non-linearities and other terms in the Hamiltonian can be added without coupling to unphysical degrees of freedom as long as the operators are expanded only as a linear combination of the physical creation and annihilation operators.

\begin{figure}
\includegraphics[trim={0cm 0cm 0cm 0cm},clip, scale=0.5]{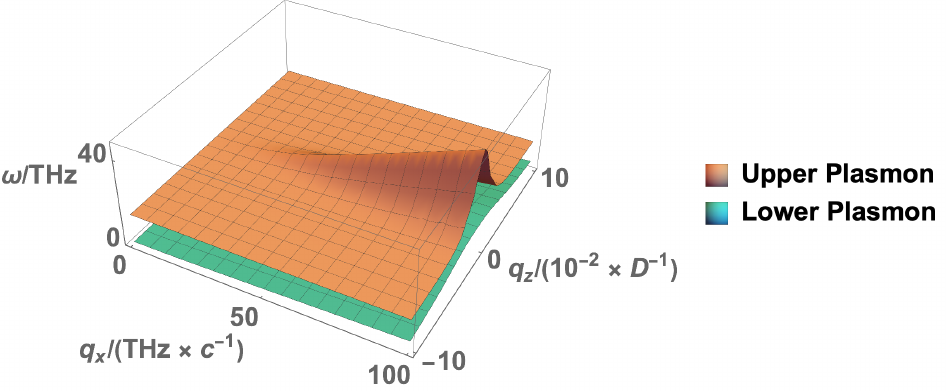}
\caption{Dispersion relation of two lower plasmons in the $\{q_x ,q_z\}$-plane. At $q_z = 0$, the upper plasmon is strongly coupled to the photon and its dispersion gradient approaches the speed of light, while away from $q_z=0$ the energy decreases.}
\label{fig:plasmons3D}
\end{figure}
\subsection{Plasmons}
\label{sec:plasmons}

Solving the equations of motion in the previous section, we find three energy scales at zero momentum which correspond to three distinct plasmons (see Appendix~\ref{app:3plasmons}). The intralayer superfluid stiffness, $\Lambda_s$, sets the energy of the in-plane plasmon which corresponds to in-plane supercurrent oscillations, 
\begin{equation}
    \omega_{\rm in-plane} = \sqrt{ \Lambda_s \frac{(e^*)^2}{ \epsilon}}
    \label{in_plane}
\end{equation}
The other two energy scales are associated with the two interlayer Josephson tunneling couplings, $\{j_{c,\lambda}\}$ and set the energy gap for the two JPs,
\begin{equation}
    \omega_{\{1,2\}} \approx \sqrt{j_{c,\{1,2\}} d_{1,2}^2 \frac{(e^*)^2}{ \epsilon}}
\end{equation}
where $\{\gamma, \epsilon,\frac{d_1}{d_2} \}$ couple the two plasmons together and offer corrections to this intuition. Due to the large separation of scale between $\Lambda_s$ and $j_{c,\lambda}$, the in-plane plasmon corresponds to a high energy excitation decoupled from the JPs that are low in energy ($\omega_{\rm upperJP} = 14.2 $ THz, $\omega_{\rm lowerJP} = 0.9 $ THz). The JPs are the relevant degrees of freedom that are accessible in the experimental set-up. The dispersion relation at small momenta is given in Fig.~\ref{fig:plasmons},\ref{fig:plasmons3D}. 
\begin{figure*}[th]
    \centering
    \includegraphics[trim={0cm 6cm 0cm 4cm},clip, scale=0.65]{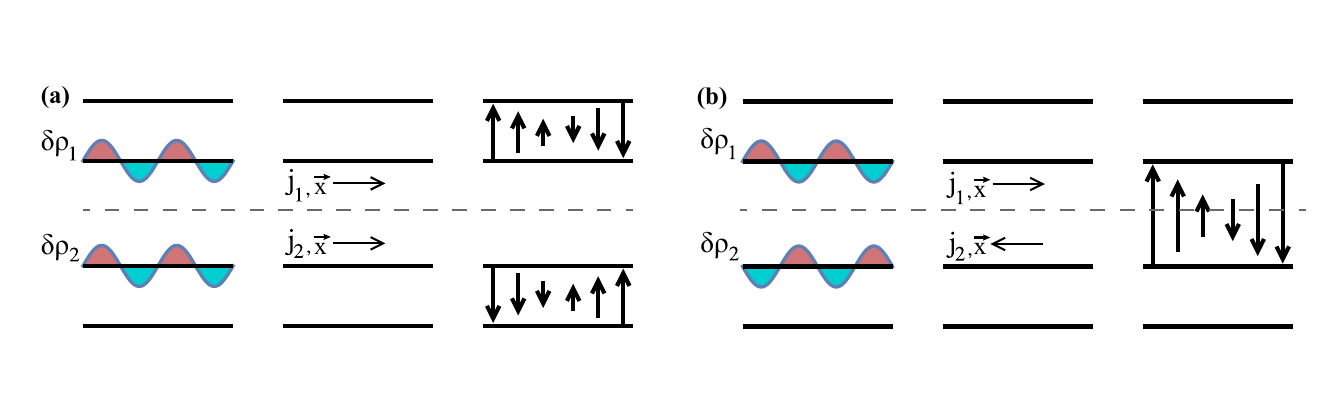}
    \caption{ Schematic representation of the pair of unstable plasmon modes, $q_z = \frac{\pi}{D}$ and $q_x$ finite. The dashed line corresponds to the crystallographic $a-b$ plane through which the reflection symmetry is defined. At this high symmetry point a) the lower plasmon is even under while b) the upper plasmon is odd.}
    \label{fig:eigenstates}
\end{figure*}

Previous studies\cite{Marel01,Bulaevskii94,Koyama96,Koyama2002,Okamoto16,Okamoto17} used the fact that the in-plane plasmon is very high in energy to essentially ignore the effects of the in-plane kinetic energy at very small momenta, $q_x << \frac{\omega_{\rm in-plane}}{c}$. While this may be sufficient to calculate optical properties of the system, we show in the next section that the phonon coupling to the in-plane kinetic energy dominates dynamics in the pump and probe experiments. Moreover, the prediction that the lower plasmon saturates at large $q_x$ is approximately true as shown by Fig.~\ref{fig:plasmons}, but fails in the non-relativistic limit, $q_x >> \frac{\omega_{\rm in-plane}}{c}$, where the in-plane kinetic energy dominates and the dispersion increases linearly with the in-plane superfluid velocity, $v_s \approx \sqrt{\Lambda_s \gamma}$ (shown in Fig.~\ref{fig:fulldisp}). 
Crucially, the presented formalism allows exploration of the full $(q_{\vec{x}}, q_z)$-plane as shown in Fig.~\ref{fig:plasmons3D}, which is necessary in order to find the most unstable mode during parametric driving (see section \ref{sec:instability}).

Eigenvectors of JPs correspond to the low energy eigenvectors of the equations of motion (\ref{eq:eigenEQ}) and allow us to project dynamics to the low energy subspace using the recipe described in Appendix~\ref{app:eigen}. In particular, this method is used in Section~\ref{sec:phonon} to derive an effective phonon-plasmon Hamiltonian starting from a physical coupling between phonons and superfluid currents. The same approach can be used to reduce non-linear corrections to the hydrodynamic equations of motion to a theory of interacting plasmons. Moreover, eigenvectors encode physical properties of JPs such as their overlap with electromagnetic fields, which are important in calculating reflectivity as outlined in subsection~\ref{sec:statrefl}.

\subsection{Reflection symmetries and collective modes in $\rm{YBa_2Cu_3O_{6+x}}$}
In the literature the phonon modes and other collective excitations are usually grouped in a convenient way with respect to their reflection symmetry in the plane parallel to the crystallographic $a$-$b$ plane. Here we comment on the symmetry of the JPs and the relevant phonon modes, and clarify what this implies for possible interactions between different modes. The Hamiltonian defined in equation (\ref{eq:hamil}) obeys the reflection symmetry in the plane parallel to the crystallographic $a$-$b$ plane, defined as:
\begin{equation}
    P \delta \rho_{\{1,2\},i}(x,t)P^\dag = \delta \rho_{\{2,1\}, - i } (x,t) 
    \label{Reflection_symmetry}
\end{equation}
with similar relations for other operators $j_{\lambda,i,\vec{x}}$, $V_{\lambda,i}$, etc. This implies that the JP eigenvectors can be written as eigenstates of the reflection operator. Written in the momentum basis, only JPs at special points in the Brillouin zone have definite reflection symmetry. Here we report that at $q_z = 0$ for arbitrary $q_x$ JPs are odd under reflection while at the edge of the Brillouin zone and at $q_z = \frac{\pi}{D}$ and arbitrary $q_x$ the upper plasmon is odd while the lower plasmon is even under reflection (shown schematically in Fig.~\ref{fig:eigenstates}). Away from these special points, linear combinations of modes with opposite momenta can be combined to produce either parity. Similarly to the electronic sector, there are two AO optical phonons of interest at zero momentum in $\rm{YBa_2Cu_3O_{6+x}}$, one at $16.5$ THz ($Q_1$) and one at $20.4$ THz ($Q_2$) (which are resonant with the probe). These photo-excited phonons are both odd under reflection symmetry. This implies that a three wave mixing between JP modes at $q_z = 0$ and a phonon is forbidden while it is allowed for finite $q_z$ and in particular at $q_z = \frac{\pi}{D}$.  As a result a three wave mixing process of the type shown in Fig.~\ref{fig:1upper1lower} involving a phonon driven parametric instability of a plasmon pair, is expected to necessarily excite plasmons at finite momentum. Such excitations can be measured in second harmonic generation experiments and the dominant pair is computed in Section~\ref{sec:instability}. On the other hand, when discussing reflectivity experiments, the incoming probe beam couples most strongly with JPs at $q_z = 0$. From symmetry arguments these modes can participate in four wave mixing processes involving two phonons and two lower plasmons at $q_z = 0$ as shown schematically in Fig.~\ref{fig:4wave}(a). The photo-induced reflectivity spectrum arising from such mixing is computed in Section~\ref{sec:refl}.

\subsection{Static reflectivity}
\label{sec:statrefl}
We conclude our discussion on the equilibrium properties of the system by computing the equilibrium reflectivity predicted by this theory. We develop a simplified analytical framework for solving the Fresnel light reflection problem for a bilayer superconducting material such as $\rm{YBa_2Cu_3O_{6+x}}$, and show that it can capture results obtained in reflectivity experiments. We will only discuss the case of normal incidence with the electric field polarized along the $c$-axis of the crystal, as shown schematically in Fig.~\ref{fig:reflStaticSchem}. We take the incident wave(signal) to be given by:
\begin{equation}
    \underline{E}_{\rm inc.} = \hat{\underline{z}} E_0 e^{ i q_x x - i \omega_s t}
\end{equation}
Correspondingly, the $B$-field lies in the $y$-direction and is given by:
\begin{equation}
    \underline{B}_{\rm inc.} = - \hat{\underline{y}} \frac{E_0}{c_{\rm vac.}} e^{ i q_x x - i \omega_s t}
\end{equation}
where $c_{\rm vac.} = \frac{1}{\sqrt{\epsilon_0 \mu_0}}$. The boundary conditions between air and material are given by the requirement that both the B-field and E-field are continuous across the boundary at every point along the interface, $x=0$:
\begin{subequations}
\begin{align}
    E_{z, \rm inc.}(z,t) + E_{z,r} (z,t) &= E_{z,t}(z,t), \\
    B_{y, \rm inc.}(z,t) + B_{y,r} (z,t) &= B_{y,t}(z,t)
\end{align}
\label{eq:boundStatic}
\end{subequations}
where $\{E_{z,r} , B_{y,r}\}$ are the amplitudes of the reflected wave and $\{E_{z,t}, B_{y,t}\}$ the transmitted wave. Due to the lattice periodic structure of the material, the equations of motion couple the $q_z = 0$ contribution of the transmitted field to all higher harmonics, $q_z = \frac{2 \pi}{D} n $, where $n$ an integer. In the spirit of the simplified model presented in Section~\ref{sec:model}, we represent the transmitted wave as piecewise constant and characterized by the interlayer fields, $\{E_{z,1}, E_{z,2} \}$. The momentum components of the electric field inside the material are given as:
\begin{figure}
    \centering
    \includegraphics[trim = {3cm 6cm 9cm 4cm},clip,scale =2]{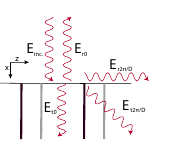}
    \caption{Schematic representation of reflection at normal incidents. Electric field is assumed to be along the $z$-axis. Due to the periodic structure of the material, light can be reflected and transmitted at momenta $q_z = \frac{2 \pi n}{D}$.}
    \label{fig:reflStaticSchem}
\end{figure}
\begin{subequations}
\begin{align}
     E_{t,q_z=0} &= \frac{ E_{z,1} d_1 + E_{z,2} d_2}{D},\\
     E_{t,q_z=\frac{2 \pi n}{D}} &= \left( E_{z,1} - E_{z,2} \right)\frac{\left( e^{-i \frac{2 \pi n}{D}}  - 1 \right)}{D}
\label{eq:qztransm}
\end{align}
\end{subequations}
and similarly for $B_{y,t,q_z}$. To satisfy the boundary conditions, light is also reflected as an evanescent wave at all these harmonics according to the boundary conditions:
\begin{subequations}
\begin{align}
    E_{z,r,q_z} &= r_{q_z} E_0, \\
    B_{y,r,q_z} &= \frac{\sqrt{\frac{\omega_s^2}{c_{\rm vac.}^2} - q_z^2}}{\omega_s} r_{q_z} E_0.
\end{align}
\end{subequations}
Note that due to the simplified piecewise structure of the transmitted fields, the boundary conditions need to be truncated to the first reciprocal lattice vector, $q_z = \frac{2 \pi}{D}$. In order to take higher harmonics into account we would need to include high energy bands, that are sensitive to the more detailed spatial structure of electric and magnetic fields between the layers. Contributions from such high energy bands are expected to be negligible at frequencies that we consider and are therefore omitted. The boundary conditions are then given by:
\begin{subequations}
\begin{align}
E_0(1+ r_0) =& E_{t,0} \\
- \frac{E_0}{c_{vac.}}(1 - r_0) =& B_{t,0}, \\
E_0 r_{\frac{2 \pi}{D}} =& E_{t, \frac{2 \pi}{D}},\\
\frac{i \sqrt{\left(\frac{2 \pi}{D}\right)^2-\frac{\omega_s^2}{c_{vac.}^2}}}{\omega_s}E_0  r_{\frac{2 \pi}{D}} =& B_{t,\frac{2 \pi}{D}}
\end{align}
\label{eq:boundstat}
\end{subequations}
where in the last line we note that $\frac{2 \pi}{D} > \frac{\omega_s}{c_{vac}}$ for the frequencies of interest and the reflected wave will be evanescent.
The transmitted electric and magnetic fields inside the material are related to each other through the equations of motion. As described in section~\ref{subsec:lorenz}, the equations of motion can be cast in the form:
\begin{equation*}
    i \partial_t \vec{v}(\vec{q}) = M(\vec{q}) \vec{v}(\vec{q})
\end{equation*}
where $\underline{v}$, is a vector containing as its elements the electromagnetic 4-component vector potential and density and phase fluctuations of the superconducting state. $\underline{\underline{M}}$ is a matrix containing gradient operators. Eigenmodes computed in section~\ref{sec:plasmons} for the bulk $\rm{YBa_2Cu_3O_{6+x}}$ material, correspond to solutions of the equations of motion of the type, $v = v(q,\omega)e^{i q x - i\omega t}$. When analyzing collective modes in the bulk, for each value of $q$, the dispersion relation of the plasmons is found by computing $\omega$. To solve the boundary problem in reflectivity experiments the problem is reversed: for each frequency, $\omega = \omega_s$, specified by the incident electromagnetic wave one computes the wavevectors, $q$, oscillating at that frequency. A challenging aspect of the light reflection problem is that one should look for solutions in the full $\underline{q}$-complex plane, where imaginary values of $\underline{q}$ are now allowed and correspond to evanescent waves, localized near the boundary. For each solution, the corresponding eigenvector encodes the relative amplitude between electric and magnetic fields for a given transmission channel. This eigenvector is found by the kernel of the matrix $\omega - M (q)$, i.e.:
\begin{equation}
   \left( \omega - M (q) \right)\cdot v_{q,\omega} = 0
\end{equation}
The electromagnetic 4-vector is expanded in the different transmission channels with complex wave-vectors $q = p + i \kappa$ as follows:
\begin{equation}
    V_{\lambda,i} = E_0 \left(t_1 v^{V_{\lambda,i}}_{1,q_1} e^{i p_1 x - \kappa_1 x} + t_2 v_{2,q_2}^{V_{\lambda,i}} e^{i p_2 x - \kappa_2 x}\right) e^{-i\omega_s t}
\label{eq:transmexp}
\end{equation}
and similarly for $\vec{A}_{\vec{x}}$ and $A_z$. In equation (\ref{eq:transmexp}), the momenta satisfying $\omega_1(q_1) = \omega_1(q_2) = \omega_s$ are the complex momentum solutions for each plasmon dispersion relation with $\kappa >0$ in our geometry, fixed by the condition that modes decay infinitely far away from the interface and $\{t_1, t_2\}$ the transmission coefficients. The electromagnetic 4-vector is converted to electric and magnetic fields through the equations 
\begin{subequations}
\begin{align}
    E_{z,\lambda,i} =& - \Delta_z V_{\lambda,i} - \partial_t A_{z,\lambda,i}, \\
    B_{z,\lambda,i} =& - \partial_x A_{z,\lambda,i} + \Delta_z A_{x,\lambda,i}.
\end{align}
\label{eq:transStatic}
\end{subequations}
Finally, substituting (\ref{eq:transStatic}) in the boundary conditions (\ref{eq:boundstat}) using equation (\ref{eq:qztransm}), we arrive at a system of four linearly coupled equations for the reflection and transmission coefficients $\{r_0, r_{\frac{2 \pi}{D}}, t_1,t_2\}$.

\begin{figure}
    \centering
    \includegraphics[scale =0.5]{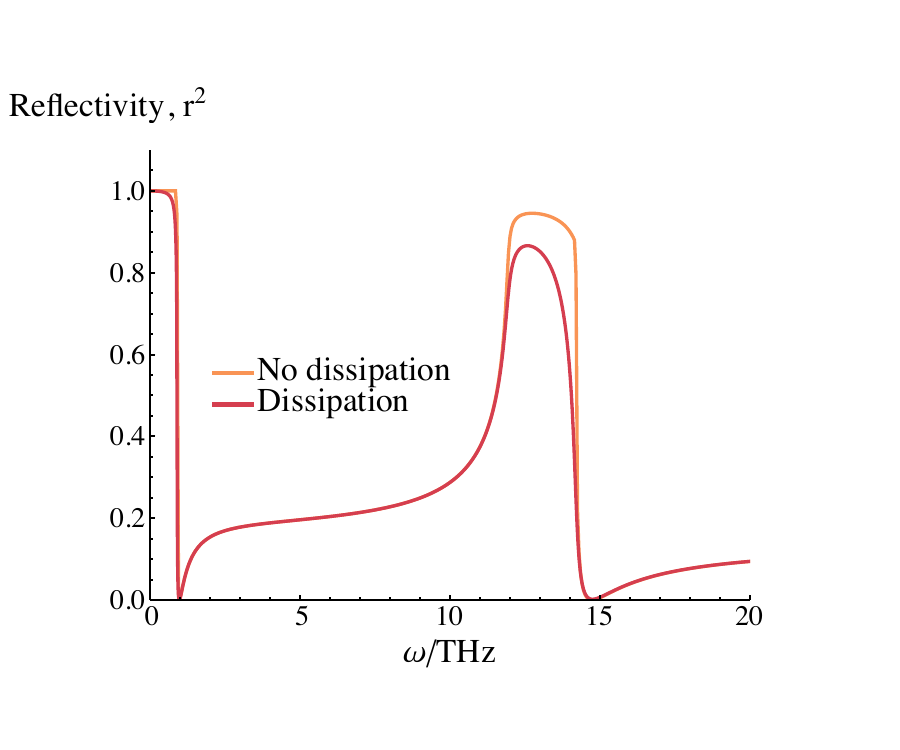}
    \caption{Reflectivity of bilayer $\rm{YBa_2Cu_3O_{6+x}}$ as a function of frequency, with and without dissipation. At frequencies above which energy transport is permitted due to the existence of a bulk collective excitation able to transmit energy inside the material, reflectivity drops abruptly from unity. Dissipation has the effect of smoothing out the features of the dissipationless state. The two edges correspond to the two JPs.}
    \label{fig:reflStatic}
\end{figure}

Using this approach, we recover the equilibrium JP edges \cite{Kaiser14} as shown in Fig.~\ref{fig:reflStatic}. The intuition behind this result comes from the fact, that at frequencies below the lower plasmon branch there are no propagating modes available to transmit energy into the material and as a result the material has reflection coefficient close to unity. As soon as a propagating mode becomes available, reflectivity drops abruptly, leading to a plasmon edge.

Quasiparticle excitations both above and below $T_c$ will lead to dissipation of collective excitations such as JPs. We take into account effects of dissipation phenomenologically by supplementing the dispersion relation with a finite negative imaginary component:
\begin{equation}
    \omega(\vec{k})^{\rm diss.}_i = \omega^{0}_i(\vec{k}) - i \gamma_i( \omega)
\label{eq:dispDiss}
\end{equation}
where $\omega^{\rm diss.}$ is the dispersion including dissipation, $\omega^{\rm 0}$ the dissipation free dispersion coming from diagonalizing the Hamiltonian and $\gamma_i >0$ is the mode dependent dissipation. Such a contribution can arise physically by including a normal current component that obeys Ohm's law. The Fresnel problem is modified to finding solutions with $\omega^{diss}_i(\vec{k}) = \omega_s$. As a result, wave-vector $\vec{k}$ can be complex even when there is a bulk propagating mode due to dissipation. Reflectivity including dissipation is shown in Fig.~\ref{fig:reflStatic}, where dissipation is seen to smooth out features of the dissipationless sate.

\section{Phonon-plasmon interaction}
\label{sec:phonon}

% {\bf I changed the order of discussion. Previously it kept jumping from one type of coupling to another and running ahead of itself. Check symmetry arguments since now they are in terms of reflection. I moved comparison of 3-wave and 4-wave processes to the next section. It was completely confusing here because it referred to proeprties of the solution of the 3-wave mixing, which readers have not seen yet. }

Following the pump pulse, $\rm{YBa_2Cu_3O_{6+x}}$ should develop a large coherent amplitude of the phonon operator at momentum $q=0$. In the presence of phonon-plasmon interactions excited phonons can provide parametric driving of JPs. Reflection symmetry discussed in Section IIB plays an important role in the analysis of phonon-plasmon coupling. 

Apical oxygen phonon is expected to modify the in-plane superfluid density and stiffness either by changing the in-plane density of carriers or by modifying their hopping. Symmetry of the phonon mode requires that these changes are antisymmetric with respect to the two layers inside one unit cell, so that $\delta \rho_{\{1,2\}}^{phon} = \pm \tilde{\xi} Q_{IR}(t) \rho_{s,0}$ and
$
     \delta \Lambda^{phon}_{s,\{1,2\}}(t) = \pm \xi Q_{IR}(t)  \Lambda_{s}
$,
where coefficients $\tilde{\xi}$ and $\xi$  characterizes the coupling strength. 

Changes in the superfluid density $\delta \rho_{\{1,2\}}^{phon}$  modify Josephson coupling between the layers, since $j_c$ should be proportional to $\sqrt{\rho_1 \rho_2}$\cite{Tinkham04}. Thus we have 
$
   \delta j_{c,\lambda}(t) = \left( \left(\tilde{\xi} Q_{IR}\right)^2  + \tilde{\xi}Q_{IR} \left( \delta \rho_2 -\delta \rho_1 \right)  \right) \frac{j_{c,\lambda}}{2\rho_{s,0}}
$ and coupling of phonons to the out-of-plane superflow kinetic energy is given by
\begin{equation}
\begin{split}
\delta H_{kin. ,z} =& \widetilde{\xi} \sum_i \int d^2 x \bigg\{ \frac{1}{2 j_{c \lambda} d^2_{\lambda} \rho_{s,0}} Q(t)^2 j_{i,\lambda,z}^2 + \\ 
&\frac{1}{2 j_{c \lambda} d^2_{\lambda} \rho_{s,0}}  Q(t)  (\delta \rho_2 - \delta \rho_1) j_{i,\lambda,z}^2 \bigg\} ,
\end{split}
\label{eq:zdirphonon}
\end{equation}
Phonon-plasmon interaction in equation (\ref{eq:zdirphonon})  describes a 4-wave mixing process. The absence of a lower 3-wave coupling in this case is a consequence of the fact that the josephson couplings between the layers, $j_{c,\lambda}$, are both even under reflection symmetry (\ref{Reflection_symmetry}). Thus, the lowest order coupling to odd phonons has to be quartic in the fields.

Phonon induced changes of the  superfluid stiffness modify in-plane superflow kinetic energy and give rise to additional phonon-plasmon interaction
\begin{eqnarray}
\delta  H_{kin,x.} = \xi \, \sum_{i}\,  \int d^2 x \, \left\{ \, \frac{Q_{IR}(t) }{2 \Lambda_s} \,    
 ( j_{1,i,\vec{x}}^2 - j_{2,i,\vec{x}}^2) \right\}
\label{eq:xdrirphonon}
\end{eqnarray}
The last equation describes 3-wave mixing interaction. Such a coupling is possible in this case since the superfluid stiffness of each layer, $\Lambda_{s,\lambda}$, neither odd nor even under reflection.
Interaction described by equation (\ref{eq:xdrirphonon}) causes a zero momentum three wave parametric process that excites pairs of plasmons at opposite momenta, a process shown schematically in Fig.~\ref{fig:1upper1lower}. Resonant processes that satisfy energy matching condition $\omega_{\rm ph.} = \omega_1(q) + \omega_2 (-q)$ lead to exponential instability discussed in Section ~\ref{sec:instability}.

It is natural to expect that lower order 3-wave processes play a greater role in phonon-plasmon interaction. We begin the next section by providing in-depth analysis of
this type of non-linearity. We comment on the effects of quartic
interactions in equation (\ref{eq:zdirphonon}) at the end of that section.

\begin{figure}
\center
\includegraphics[trim={0cm 0cm 0cm 0cm},clip, scale=0.5]{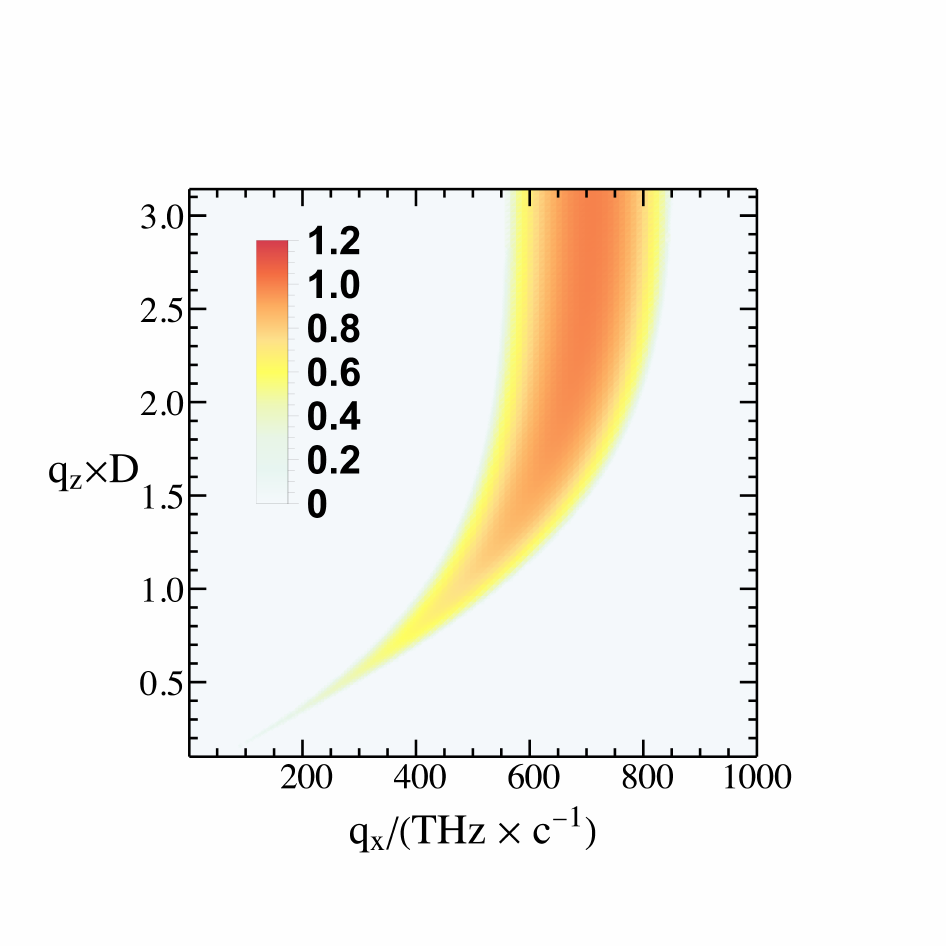}
\caption{ Relative growth rate of unstable modes in $\{q_x,q_z\}$-plane. The relative growth rate is measured in $THz$ while to obtain the true growth rate this result needs to be multiplied by the relative change in the superfluid density caused by the phonon amplitude. The coupling of phonons to the in-plane kinetic term grows quadratically with the in-plane momentum, an observation reflected by larger growth rate for instabilities with larger in-plane momentum. The most unstable JP pair is made up of a lower plasmon at $2.2$ THz and an upper plasmon at $14.3$ THz}
\label{fig:intersec}
\end{figure}

\section{Instability Analysis}
\label{sec:instability}

In this section we analyze the dynamical stage of the pump and probe experiments in which coherently excited phonons give rise to parametric instability of plasmons. Short-time dynamics can be deduced by treating the phonon mode as an external classical field, ignoring back action coming from plasmons.
% This allows us to perform an instability analysis of the equilibrium state in the presence %of the phonon drive. 
From such an analysis, we seek to find the fastest growing mode which is expected to dominate the dynamics. We project dynamics to the low energy sector corresponding to JP relevant to parametric resonant processes involving the AO phonon, by expanding the original variables in the eigenbasis of the equilibrium Hamiltonian, Eq.~(\ref{eq:hamil}), using the prescription in equation~(\ref{eq:eigenDec}) and keeping only the JPs' creation and annihilation operators. The phonon driven plasmon Hamiltonian can then be written as 
\begin{equation}
\begin{split}
H =& \sum_q \sum_{i=1,2} \omega_i(q) b_{i}^\dag(q) b_{i}(q) + \\
&\sum_q \sum_{i,j = {1,2}} g_{i,j}(t,q) b^\dag_i(q) b^\dag_j(-q) + h.c. , \\
\mbox{\qquad}&
\label{eq:paraDriven}
\end{split}
\end{equation} 
where the index $i$ corresponds to the two plasmon branches and not a unit cell index. In equation~(\ref{eq:paraDriven}), $\omega_i(q)$ is the dispersion relation of the two plasmons, while the time-dependent squeezing term, $g_{i,j}(q,t) b_i^\dag(q) b_j^\dag(q) $, arises from the parametric drive in equation~(\ref{eq:xdrirphonon}) and is responsible for the exponential instability on parametric resonance. The amplitudes, $g_{i,j}(q,t) = \xi Q_0 \sin(\omega_{\rm ph} t)  g_{i,j}(k) $, are found from the eigenvector expansion. We simplify the discussion by considering only interband coupling, since intraband process are far detuned from resonance. This amounts to setting $g_{1,1} = g_{2,2} = 0$ and $g_{1,2}(q) = g$ where for brevity the momentum dependence has been suppressed but it is always implicit. The time-dependent Hamiltonian in equation~(\ref{eq:paraDriven}) can be turned into a time-independent infinite dimensional banded matrix by transforming to the Floquet basis:
\begin{equation}
    b_{i}(t) = \int_{0}^{\omega_{\rm ph}} d \omega \sum_n e^{ - i n \omega_{\rm ph} t - i \omega t} b_{i,n} ( \omega ) 
\label{eq:floquetAnsatz}
\end{equation}
where $n$ is the Floquet band index. In this basis the drive appears as coupling between different Floquet bands, and the equations of motion take the form:
\begin{widetext}
\begin{equation}
i \omega \begin{pmatrix} 
b^\dag_{1,n+1} (\omega) \\
b_{2,n} (\omega )
\end{pmatrix} = \begin{pmatrix} i\left( \omega_1 - (n + 1)\omega_{ph} \right) && i g^* \xi Q_0 \\
- i g \xi Q_0 && - i \omega_2 - n \omega_{\rm ph}
\end{pmatrix} \begin{pmatrix} 
b_{1,n+1}^\dag (\omega) \\
b_{2,n}(\omega)
\end{pmatrix},
\label{eq:FloquetEOM}
\end{equation}
\end{widetext}
The above $2\times2$ matrix equation represents an infinite dimensional matrix that couples all Floquet sectors to each other. To find Floquet eigenstates one needs to diagonalize equation~(\ref{eq:FloquetEOM}). Instabilities correspond to eigenvalues with imaginary frequency components signaling exponential growth. To find the growth rate of instabilities to leading order in the drive's strength it is enough to diagonalize only two nearest neighbour Floquet bands, an approximation called Floquet degenerate perturbation theory and is equivalent to a first order Magnus expansion \cite{Eckardt15}.  Here we sketch arguments for the validity of this approximation. Condition for parametric resonance is given by
\begin{equation}
\omega_{\rm ph} = \omega_1 + \omega_2.
\end{equation}
This condition makes the two different states at neighbouring Floquet bands degenerate. As a result, degenerate perturbation theory can be used so that the leading contribution to the eigenvalues is given by diagonalizing the matrix only in the degenerate subspace. Note that for each $n$ there is an equivalent, decoupled, degenerate subspace and hence we can set w.l.o.g. $n = 0$. On parametric resonance the eigenvalues of the above problem are found to be
\begin{equation}
    \omega = - \omega_{\{1,2\}} \pm i \left|g  \xi Q_0\right|
\label{eq:floquetEigen}
\end{equation}
Perturbative corrections in powers of $Q_0$ can be found by including more Floquet bands. From equation~(\ref{eq:floquetEigen}), we deduce that the growth rate is given directly by the matrix element $g \xi Q_0$. Away from resonance, parametrizing the detuning by $\omega_{\rm ph} = \omega_1 + \omega_2 + \delta$, the growth rate is given by
\begin{equation}
\omega= -\left( \omega_{\{1,2\}}(q) + \frac{\delta}{2}\right) \pm i \sqrt{ \left| g(q) \xi Q_0 \right|^2 - \left(\frac{\delta}{2}\right)^2}
\label{eq:floquetDet}
\end{equation}
where the momentum dependence has been restored for concreteness. The detuning, $\delta$, suppresses the growth rate. The leading instability is given by a plasmon pair at the wave-vector which maximizes equation~(\ref{eq:floquetDet}). Such a mode is expected to grow exponentially faster than every other mode and therefore dominate the dynamics. The relative growth rate as a function of momentum is plotted in Fig.~\ref{fig:intersec}. The central trace corresponds to the instability condition curve $\omega_{\rm ph} = \omega_1(q) + \omega_2 (q)$. As argued in Section~\ref{sec:phonon}, the dominant parametric driving term comes from equation~(\ref{eq:xdrirphonon}). This implies that the coupling $g(q)$ is expected to grow as $q_x^2$ while depending only weakly on $q_z$. Our result confirms this intuition, with the unstable mode with largest $q_x$ growing the fastest. Surprisingly, this occurs at the edge of the Brillouin zone along the $c$-axis at $q_z = \frac{\pi}{D}$. The fact that the most unstable mode occurs at the largest possible $q_z$ is expected to be a prediction robust to details.

\begin{samepage}
Before concluding this section we discuss our justification for not including quartic phonon-plasmon coupling described by equation~(\ref{eq:zdirphonon}).
The first term in equation~(\ref{eq:zdirphonon}) can be treated in the same way as the three wave mixing process. Since it is quadratic in the plasmons fields it can still be thought of as parametric driving of plasmons by an external field, which in this case is given by $Q^2(t) \rightarrow \expe{Q(t)}^2$. This process has a different resonant condition and when the phonon amplitude  $\expe{Q(t)}$ is small, we expect it to be subdominant. In the next section we point out, however, that this mechanism may play an important role in the renormalization of TerHertz reflectivity. Another possibility is a quartic term that has one phonon and three plasmons fields. It can be studied by a different type of instability analysis (see Appendix~\ref{app:flucres}). Even though such a term can still lead to an instability, upon Fourier transforming this term in the momentum basis, which is the energy eigenbasis, one notes that the amplitude of the interaction is suppressed by an overall $\frac{1}{\sqrt{V}}$, where $V$ is the system's size, making this interaction irrelevant.
\end{samepage}

%%%%%%%%%%%%%%%%%%%%%%%%%%%%%%%%%%%%%%%%%%%%%%%%%%%%%%%%%%%%%%%%%%%%%%%%%%%%%%%%%%%%

\section{Reflectivity of Floquet Medium}
\label{sec:refl}

\begin{figure}
    \includegraphics[trim={3cm 4.5cm 5cm 7cm},clip, scale=0.9]{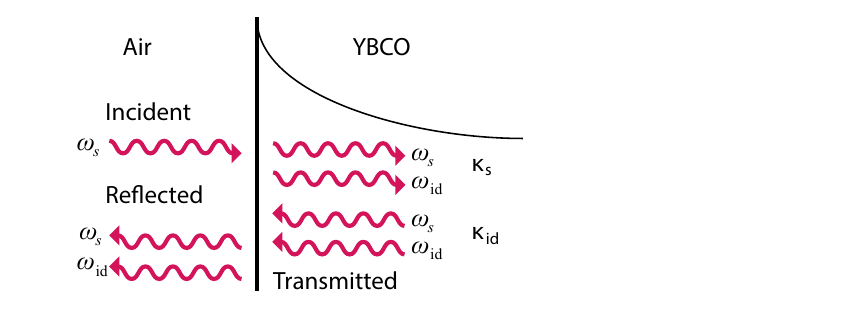}
    \caption{Schematic reflectivity from a Floquet medium. Oscillating fields inside the medium induce frequency mixing between incident wave frequency $\omega_s$ and idler $\omega_{id}$. In a Floquet medium, transmission solutions appear that propagate towards the surface amplifying reflectivity. }
\label{fig:floquetenhancement}
\end{figure}

In the previous section we discussed a process in which a strong coherent excitation of AO phonons leads to parametric instability of JPs. Such an instability populates classically the most unstable JP modes generating supercurrents oscillating at the frequencies of these modes. Such currents have been observed by von Hoegen et al\cite{vanHoegen19} using the optical second harmonic generation technique.
In this section we analyze TeraHertz reflectivity 
of the non-equilibrium state of $\rm{YBa_2Cu_3O_{6+x}}$ with strongly excited phonons and Josephson plasmons. We will demonstrate that resonant parametric mode coupling naturally explains a variety of experimental signatures observed by Hu et al.\cite{Hu14} and Kaiser et al.\cite{Kaiser14}. As mentioned in the introduction, non-trivial signatures of the phonon pump have been found both below and above $T_c$.  Below $T_c$, the appearance of a photo-induced peak close to the lower JP edge and red-shifting of the upper JP egde. Above $T_c$, the re-emergence of the lower JP edge at a blue shifted frequency and the re-emergence of the upper JP edge at a red-shifted frequency from a featureless equilibrium spectrum in the absence of a pump.

To simplify theoretical analysis, we treat the oscillating fields as monochromatic drives that parametrically modify the equilibrium JP modes through phonon-plasmon or plasmon-plasmon non-linearities. This assumption is well justified on time scales shorter than the decay time of the oscillating modes. At longer times decay processes can be included by considering drive fields to have a range of frequencies. We postpone the discussion of the effects of the  driving fields depletion until future publications. 
We consider the following effective model for the driven Josephson Plasmons
\begin{equation}
H_{\rm eff,int} = \sum_{\alpha}\sum_{q} \sum_{i,j = 1,2} g_{i,j; \alpha} e^{ i \Omega_{\rm fl,\alpha} t} b^\dag_{i} (q) b_j^\dag(-q + Q_{\alpha}) + h.c. 
\label{eq:hamileffFloq}
\end{equation}
where $\alpha$ is the index of a particular oscillating field, while $\Omega_{\rm fl,\alpha}$ and $Q_{\alpha}$ are the frequency and momentum of the oscillating fields respectively. Another underlying assumption of (\ref{eq:hamileffFloq}) is retaining only the pair creation processes and neglecting re-scattering terms of form $e^{i \Omega_{\rm fl,\alpha} t} b^\dag_{i} (q) b_j(q - Q_{\alpha})$. The latter are not expected to play an important role in the mechanism of parametric driving. 

When analyzing TeraHertz reflectivity of the non-equilibrium system described by equation (\ref{eq:hamileffFloq}) we assume that for every frequency window of interest it is sufficient to include only one most relevant driving field. This corresponds to considering only the process whose frequency is closest to the parametric resonance condition. Effects coming from resonant process are expected to strongly dominate over the non-resonant ones\cite{offresonant20,Rodriguez18}. Finally we note that symmetries impose further constraints on the allowed non-linear processes. Here we discuss reflection at normal angle of incidence and therefore only need to consider modes at $q_z = 0$. An important manifestation of symmetry constraint is that the dominant three-wave mixing process that gives rise to the parametric instability of JPs (see discussion in the previous section) does not contribute to reflectivity renormalization directly: this coupling vanishes for $q_z=0$ because all three modes are odd under the reflection symmetry discussed in Section IIC. Another
nontrivial consequence of the symmetry is that dominant non-linear process affecting modes close to the lower JP edge will be different compared to the one renormalizing the upper JP edge (see discussion below).

When discussing light interacting with a Floquet medium it is common to define the $signal$ and $idler$ frequency components where $\omega_s$ is the frequency of the probe and $\omega_{id} = \Omega_{\rm fl} - \omega_s$( shown schematically in Fig.~\ref{fig:floquetenhancement}). Within Floquet perturbation theory it is sufficient to consider mixing between these two modes and neglect the admixture of higher Floquet harmonics. To compute reflectivity, we extended 
the theoretical framework for solving the Fresnel problem introduced in Subsec.~\ref{sec:statrefl} for the static case to the case of a Floquet medium. Details are presented in Appendix~\ref{app:floqfresn}.

We identify three features that are expected to be ubiquitous in the reflectivity spectra of all Floquet systems and find that each experimental feature mentioned above originates from a combination of these three effects. These effects are:
\begin{itemize} 
\item \textit{Parametric amplification on resonance:} Probing a Floquet medium at frequencies close to parametric resonance (i.e. at half the frequency of the driving field) leads to an enhancement of reflectivity and appears as a photo-induced peak.  
\item \textit{Modes close to parametric resonance become more coherent:} The exponential growth caused by the drive on parametric resonance compensates for the decay due to dissipation.
\item \textit{Away from parametric resonance modes are renormalized corresponding to the Floquet level attraction:} As shown in Appendix~\ref{app:dressed}, away from resonance, states coupled through the drive are dressed by the drive. The dressed energies are shifted compared to equilibrium according to a simple level attraction rule.
\end{itemize}
We now proceed to explain each of the experimental features mentioned above.

\begin{figure}
    \centering
    \includegraphics[scale = 0.5]{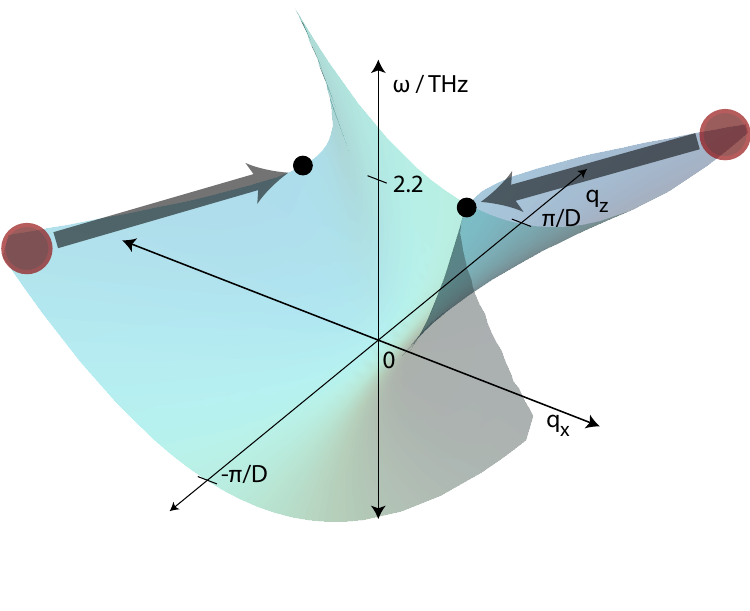}
    \caption{Schematic representation of the possible JP four wave mixing resonant process. A pair of parametrically unstable JPs at $q_z = \frac{\pi}{D}$ and $\omega = 2.2$ THz can decay into a pair of JPs at $q_z = 0$, also at $\omega = 2.2$ THz}
    \label{fig:4waveplasmons}
\end{figure}

\begin{figure}
    
    \includegraphics[ scale=0.6]{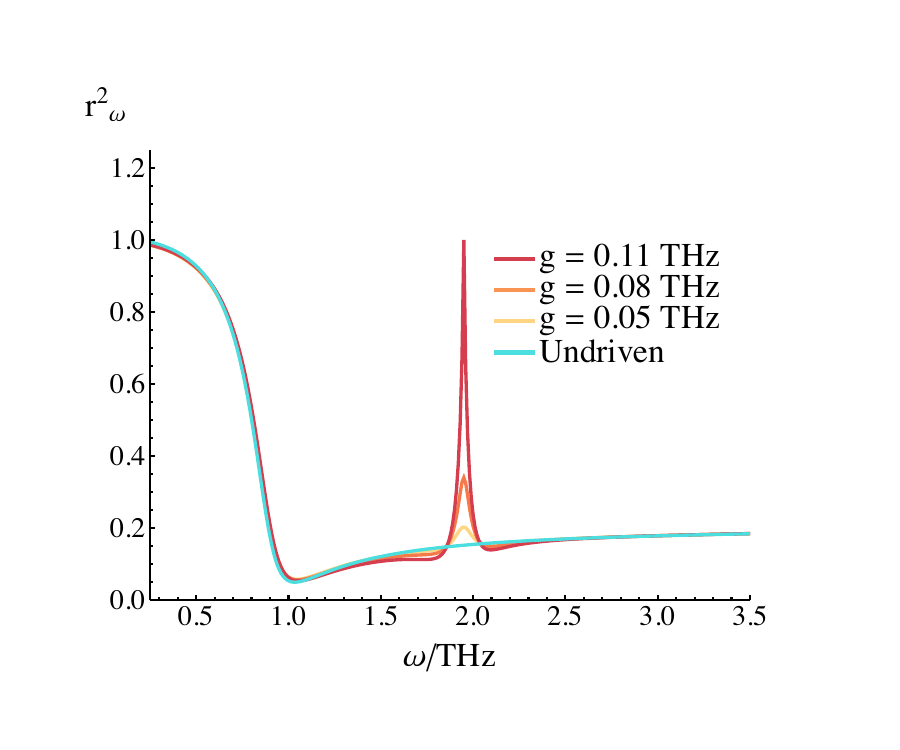}
    \caption{Reflectivity at the probe frequency as a function of frequency for Floquet $\rm{YBa_2Cu_3O_{6+x}}$ after the pump in the regime of small dissipation. The leading oscillating term considered is a two phonon four wave mixing process oscillating at the difference of two adjacent phonon modes at $\Omega_{\rm fl} = 3.9$ THz. As the amplitude of oscillation is increased, a photo-induced peak appears on parametric resonance, $\omega = 2$ THz. Another possible mechanism of this peak is four plasmon resonant scattering shown in figure \ref{fig:4waveplasmons}.}
\label{fig:T0floqrefl}
\end{figure}

\paragraph{Photo-induced peak in reflectivity @ $1.8 - 2.4$ THz below $T_c$.} At normal incidence, the transmitted wave is at $q_z = 0$. As mentioned in Section~\ref{sec:model}, both lower and upper plasmon at $q_z = 0$ are odd under parity. This implies that the dominant three wave mixing process that gave rise to the parametric instability of JPs can not contribute to the reflectivity at normal incidence. On the other hand, the four wave mixing process ignored in the last section can now contribute and corresponds to a parametric drive with amplitude $\expe{Q_1(t)} \expe{Q_2(t)}$, shown schematically in Fig.~\ref{fig:4wave}(a). Such a drive oscillates at a frequency corresponding to the difference between phonon $Q_1$ at $\sim 16.5$ THz and the adjacent phonon $Q_2$ at $\sim 20.4$ THz which is also excited during the pump process as reported in experiments of Hu et al.\cite{Hu14} and Kaiser et al.\cite{Kaiser14}. This leads to a drive at zero momentum and frequency $\omega_d = 3.9$ THz. Another possible mechanism is a four wave mixing process between JPs shown in Fig.~\ref{fig:4waveplasmons}. JPs at frequency $2.2$ THz and $q_z = \frac{\pi}{D}$, excited by AO phonons through the parametric resonance process mentioned in the previous section, can themselves become a source of parametric driving. For example, they can drive the lower JPs at $q_z = 0$, which play the key role in light reflection near the lower JP edge. A similar mechanism has been observed by Baumberg et al. for excitons  near the so called magic angle \cite{Savidis1996}. Both mechanisms provide a parametric drive at $q = 0$ but at slightly different frequencies ($3.9$ THz for the phonon mechanism and $4.4$ THz for the JP mechanism). As we will discuss below, both scenarios produce similar results and are consistent with experimental data. We focus here on the phonon four-wave mixing mechanism. In this scenario, the effective interacting Hamiltonian keeping only resonant time-dependent contributions is given by:
\begin{equation}
    H_{eff.,int} = \sum_q  g^{lower}_{q} \expe{Q_1(t)} \expe{Q_2(t)} b_1^\dag(q) b_1^\dag(-q)
\label{eq:effLower}
\end{equation}
In the absence of dissipation, reflectivity as a function of frequency in the presence of this phonon four wave mixing term is plotted in Fig.~\ref{fig:T0floqrefl}. The most striking feature is the appearance of a photo-induced peak at the parametrically resonant frequency $\frac{\omega_d}{2} = 2$ THz. This suggests that the experimental feature of a photo-induced peak arises from parametric amplification of reflectivity due to the four wave phonon-plasmon process. 

\begin{figure}
    \includegraphics[ scale=0.6]{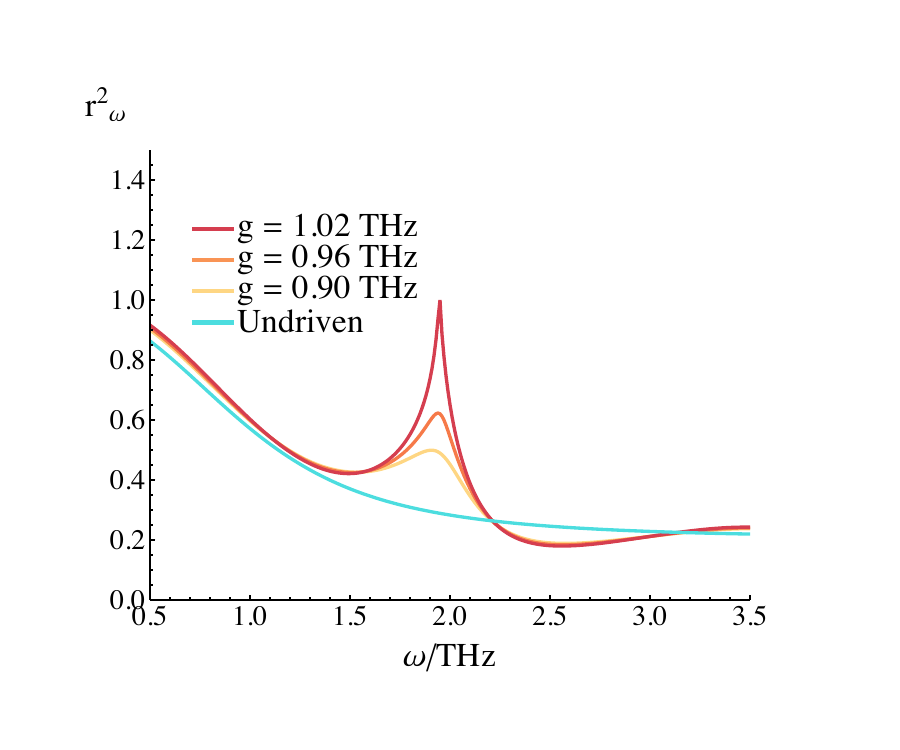}
    \caption{Reflectivity of Floquet medium in the overdamped case. Parametric amplification of the probe beam at $2$ THz re-establishes the lower plasmon edge at a blue-shifted frequency, $\omega = 2$ THz.}
    \label{fig:TinffloqRefl}
\end{figure}

\paragraph{Re-emergence of lower JP edge @ $1.8 - 2.4$ THz above $T_c$.}We now address the question of the emergence of the lower JP edge in the pseudogap regime reported by Hu et al.\cite{Hu14} and Kaiser et al.\cite{Kaiser14}. We assume that at equilibrium in this regime, JPs can be described as overdamped modes which do not produce a plasmon edge. The situation for strong dissipation, $\gamma_1 = 1$ THz (see Eq.~\ref{eq:dispDiss}), is shown in Fig.~\ref{fig:TinffloqRefl}. In the absence of a drive, reflectivity is featureless. However, for strong enough drive the plasmon edge re-emerges at the blue shifted frequency around $\frac{\omega_d}{2}$ where parametric driving compensates dissipation most efficiently reviving features of the dissipationless state. 

In the case of the JP four-wave mixing mechanism shown in Fig.~\ref{fig:4waveplasmons}, the results would be the same but with a slightly shifted parametric resonance frequency at $\frac{\omega_d}{2} = 2.2$ THz. The experimental results place the above features at $1.8 - 2.4$ THz, thus both scenarios are consistent with these experiments. One way to determine which mechanism dominates is to pump the system with a narrow bandwidth pulse. If only one of the phonons is excited but the feature remains, this would be evidence that the JP four wave mixing process is responsible for this reflectivity feature instead of the phonon four-wave mixing process.

\paragraph{Red-shift of upper JP edge below $T_c$.}
\begin{figure}
    \includegraphics[trim={4cm 0.6cm 2.5cm 0.6cm},clip, scale=0.4]{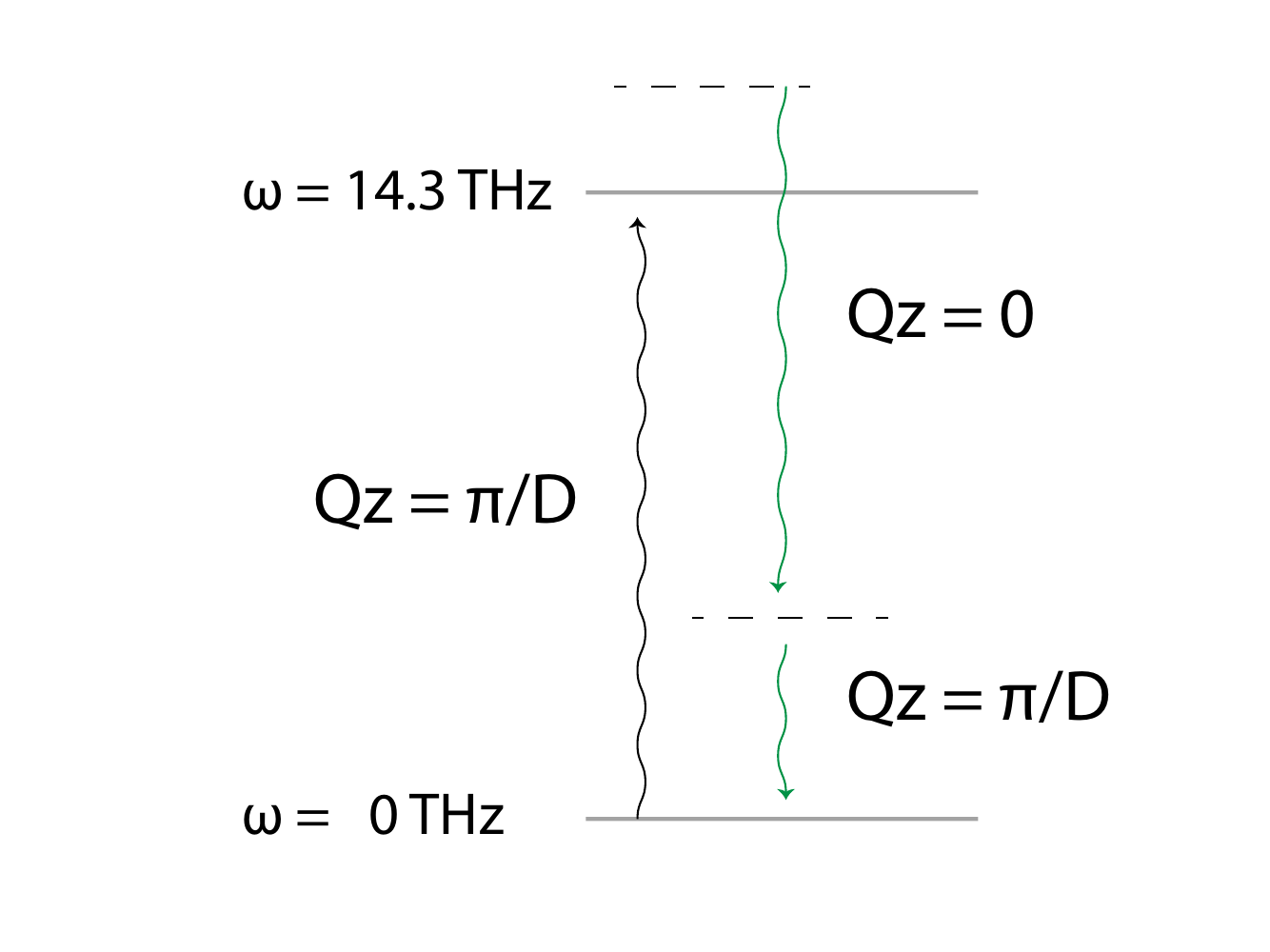}
    \caption{Schematic representation of the three wave mixing process renormalizing the upper plasmon edge. The instability induced pump corresponding to the expectation value of the upper plasmon at $Q_z = \frac{\pi}{D}$, $Q_x = 500$ THz/c and $\omega = 14.3$ THz acts as an external drive (black). This drive couples the upper plasmon band at $Q_z = 0 $ (green) to the lower plasmon band at $Q_z = \frac{\pi}{D}$ (green). However, the sum of the energies of the upper and lower plasmon are higher than $\omega = 14.3$ THz which leads to a red-shift of the upper plasmon.}
    \label{fig:3waveupper}
\end{figure}

\begin{figure}
    
    \includegraphics[ scale=0.6]{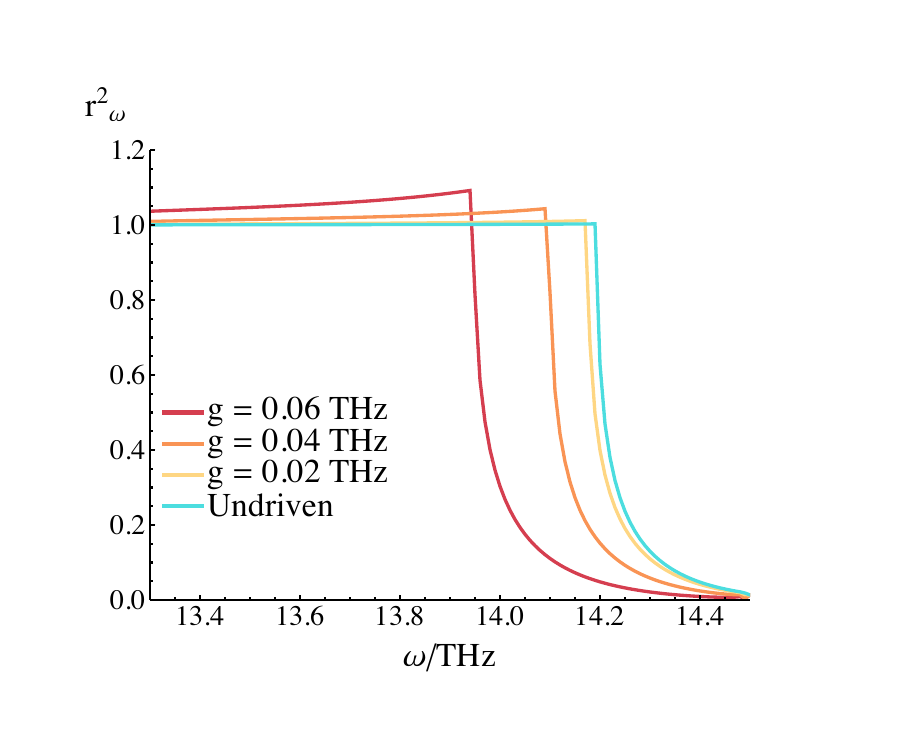}
    \caption{Reflectivity excluding the idler component as
a function of frequency of the Floquet medium with weak dissipation, with $\Omega_{\rm fl} = 14.3$ THz at different driving
amplitudes. The upper plasmon edge is red shifted.}
\label{fig:T0floqrefl14}
\end{figure}

At frequencies around $14$ THz corresponding to the upper plasmon edge, the four-wave mixing term involving two phonons mentioned in the previous section is far-detuned. In this frequency window, the resonant process that obeys the parity symmetry constraints is a three-wave mixing process between JPs. This process involves the upper plasmon at $Q_z = \frac{\pi}{D}$ and $Q_x = 500$ THz/c and oscillating at $14.3$ THz converted into an upper plasmon at $q_z = 0$ and a lower plasmon at $q_z = \frac{\pi}{D}$ to conserve momentum, shown schematically in Fig.~\ref{fig:3waveupper}. The upper plasmon at $Q_z = \frac{\pi}{D}$ and $Q_x = 500$ THz/c corresponds to an unstable mode which is classically populated and is treated as an external drive. In this scenario, the effective interacting Hamiltonian takes the form:
\begin{equation}
\begin{split}
    H_{eff.,int} =& \sum_{q_x}g^{upper}_{q} \expe{b_2(Q_z, Q_x)} \\ &\times  b_2^\dag(q_z = 0 ,q_x) b_1^\dag(Q_z, Q_x - q_x)
\end{split}
\label{eq:effUpper}
\end{equation}
In equation~(\ref{eq:effUpper}), we consider the upper plasmon mode, $b_2(q_z = 0, q_x)$, which couples to the incoming probe beam at normal incidence whose momentum along the c-axis is fixed to $q_z = 0$ from boundary conditions. The upper plasmon at $Q = \{Q_z, Q_x\}$ has a classical expectation value and acts as an external drive. Finally, to preserve both momentum conservation and parity conservation the mode coupled to the probe beam is dressed through the drive with the state $b_1^\dag(Q_z, Q_x -q_x)$ renormalizing its energy. Level attraction (Appendix~\ref{app:dressed}) implies that this parametric driving process will result in a red shift of the upper plasmon edge as the sum of the bottom of the two bands ($0.9$ THz + $14.2$ THz) is greater than the frequency of the drive ($14.3$ THz). This situation is demonstrated in Fig.~\ref{fig:T0floqrefl14}. 

\begin{figure}
    
    \includegraphics[ scale=0.6]{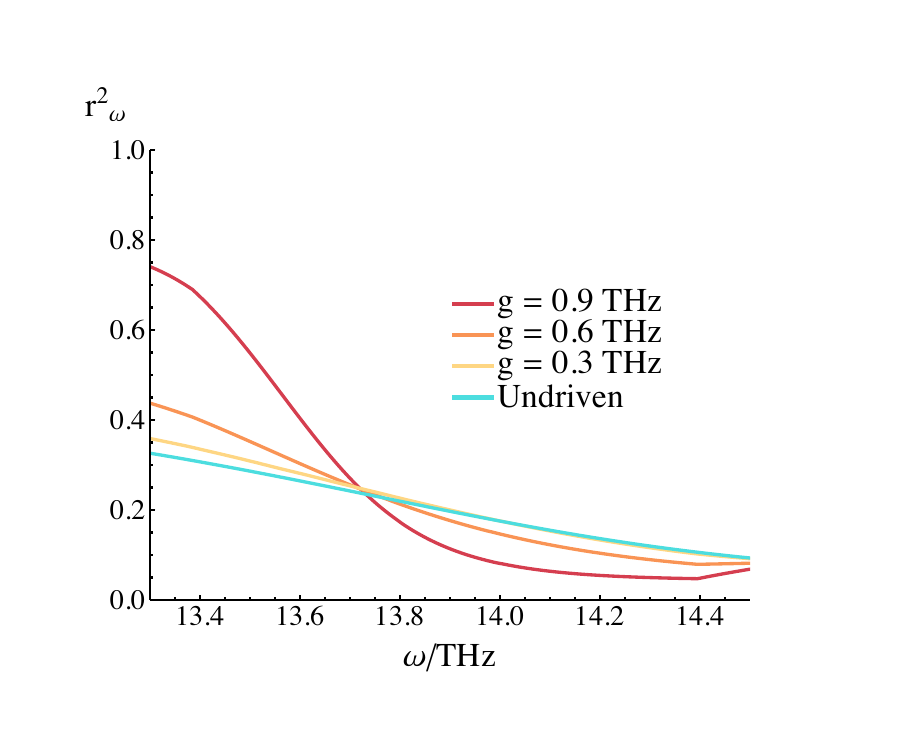}
    \caption{Reflectivity at probe frequency near the upper plasmon edge as a function of frequency in the presence of dissipation. Parametric driving from the upper plasmon induced pump revive the upper plasmon edge at a red shifted value.}
\label{fig:Tinffloqrefl14}
\end{figure}

\paragraph{Re-emergence of upper JP edge at red-shifted frequency above $T_c$.} The pseudogap phase is modelled  phenomenologically through strong dissipation as in the previous discussion. The reflectivity in the pumped state with strong dissipation is plotted in Fig.~\ref{fig:Tinffloqrefl14} and shows clear indication of a re-emergence of the upper JP edge at a red-shifted frequency in agreement with the experiments. This implies that this experimental signature is the result of parametric driving making equilibirium modes more coherent close to resonance and the energy shift caused by level attraction. 

\section{Conclusions and Outlook} 
\label{sec:concl}

In summary, we developed a unified theoretical model that describes pump and probe experiments in $\rm{YBa_2Cu_3O_{6+x}}$, in which resonant excitation of apical oxygen phonons  lead to dramatic changes in the low frequency reflectivity both below the superconducting $T_c$ and in the pseudogap phase. Our analysis provides a general framework for understanding  generation of currents following a pump resonant to the AO phonons, as well as re-emergence of the lower plasmon edge above $T_c$ at blue shifted frequency and of the upper plasmon edge at a red shifted frequency. Finally, we were able to explain the appearance of a photo-induced peak in the reflectivity below $T_c$ and the renormalization of the upper plasmon edge to lower frequencies. We demonstrated that the origin of all these phenomena is in the non-linear couplings between phonons and Josephson plasmons, with the first step being parametric excitation of pairs of Josephson plasmons by strongly photo-excited AO phonons. The latter process originates  from a renormalization of the in-plane superfluid density due to the apical oxygen's motion; for such a coupling the parametric resonance condition can be satisfied for pairs of plasmons in the upper and lower branches at opposite momenta. This analysis allowed us to compute the growth rate of JP pairs in the full $q_\parallel ,q_z$ - plane.  Our results identify an instability curve in the momentum plane for the unstable plasmons which lies outside the light cone. Subsequently, we proceeded to calculate reflectivity in equilibrium and in the pumped state. In equilibrium, we developed a framework to solve the Fresnel problem in a bilayer systems such as $\rm{YBa_2Cu_3O_{6+x}}$. This framework was extended to a Fresnel-Floquet problem which takes into account the effects of an oscillating field using Floquet degenerate perturbation theory.

We suggested that the pseudogap phase can be understood as hosting overdamped plasmons, and discussed that this scenario provides a good description for the observed phenomena of photo-induced superconductivity in $\rm{YBa_2Cu_3O_{6+x}}$. Going beyond phenomenological description of this regime is beyond the scope of the current paper. Such analysis can be done by extending earlier work on Josephson plasmons in the vortex liquid phase \cite{Bulaevskii1992,Kleiner94,Bulaevskii96,Koshelev1996,Koshelev2000,Koshelev2014}.  Thermal fluctuations of the pairing amplitude are expected to lead to plasmon broadening and suppression of the parametric instability \cite{Koshelev2014}. Hence detailed analysis of phonon-plasmon three-wave parametric resonance  above Tc should provide valuable insight into the nature of superconducting fluctuations in the pseudogap regime \cite{Uemura1989,Emery95,Paramekanti2000,Benfatto2001}.
We plan to present such an analysis in subsequent publications.

Before concluding this paper, we would like to point out that plasmons generated from photoexcited phonons through parametric instability do not radiate out of the sample because of the mismatch of energy and momentum with electromagnetic waves in vacuum. However they can be made to radiate into the far field, if one adds a meta-material type structure on the surface of a sample, which compensates for the momentum mismatch (see e.g. structures in references\cite{Meng_20,Schalch19}). By varying the spatial periodicity of such structures, one should be able to enhance emission of Josephson plasmons at different momenta. This opens  promising new direction for generation of Terahertz waves as entangled
pairs with tunable frequencies. 

\section*{Acknowledgments}

We are grateful for useful discussions with Pavel Dolgirev, Kushal Seetharam, Leonid Glazman, Israel Klich, Patrick Lee and Bertrand I. Halperin, Daniel Podolsky, Sho Sugiura, Joe Orenstein, Matteo Mitrano, Dieter Jaksch, Rick Averitt, Andy Millis and Antoine Georges. M.H.M. and E.D. acknowledge support from Harvard-MIT CUA, AFOSR-MURI: Photonic Quantum Matter (award FA95501610323),
DARPA DRINQS program (award D18AC00014). A.vH., M.F, M.F and A.C. acknowledge funding from the European Research Council under the European Union's Seventh Framework Programme (FP7/2007- 2013)/ERC Grant Agreement No. 319286 (Q-MAC). A.C. acknowledges support from the Deutsche Forschungsgemeinschaft (DFG) via the Cluster of Excellence 'The Hamburg Centre for Ultrafast Imaging' (EXC 1074 - Project ID 194651731) and from the priority program SFB925.

\appendix

\section{Mathematical details of diagonalization procedure}
\label{app:eigen}

In this paper we use a generalized Bogoliubov transformation in order to transform the Hamiltonian in Eq.~\ref{eq:hamil} written in terms canonically conjugate pairs with generalized coordinates, $\vec{q}_i = \{\rho_{i,\lambda}, V_{i,\lambda}, \vec{A}_{\vec{x}}, A_{z,i ,\lambda} \}$, and conjugate momenta $\vec{p}_i = \{\phi_{i,\lambda}, P_{V_{i,\lambda}}, P_{\vec{A}_{\vec{x}}}, P_{A_{z,i ,\lambda}} \}$, to a diagonal Hamiltonian written in terms of creation and annihilation operators for the elementary excitations of the system (plasmons). Our notation follows closely Ref.~\cite{xiao09theory}. We define the spatial Fourier transform as:
\begin{subequations}
\begin{align}
    q_i(\vec{x})=&    \int_{-\frac{\pi}{D}}^{
    \frac{\pi}{D}}\frac{d k_z}{\sqrt{2 \pi}} \int \frac{d^2 k_x}{2\pi} e^{i \vec{k}_x \vec{x} + i k_z D i} q (\vec{k}) ,\\
     p_i(\vec{x})=&   \int_{-\frac{\pi}{D}}^{
    \frac{\pi}{D}}\frac{d k_z}{\sqrt{2 \pi}} \int \frac{d^2 k_x}{2\pi} e^{i \vec{k}_x \vec{x} + i k_z D i} p (\vec{k}).
\end{align}
\end{subequations}
The Hamiltonian is translationally invariant and therefore diagonal in momentum space and takes the form:
\begin{equation}
    H = \int_{-\frac{\pi}{D}}^{
    \frac{\pi}{D}}\frac{d k_z}{\sqrt{2 \pi}} \int \frac{d^2 k_x}{2\pi} \left(\frac{1}{2} \phi^\dag(\vec{k}) \cdot A(k) \cdot \phi(\vec{k}) \right)
\label{eq:hamilMom}
\end{equation}
where $A$ is a Hermitian matrix. In equation (\ref{eq:hamilMom}), the vectors $\phi^\dag(\vec{k})$ and $\phi(\vec{k})$, are defined as
\begin{equation}
    \phi(\vec{k}) = \begin{pmatrix} \vec{q}(\vec{k})\\ \vec{p}(\vec{k}) \end{pmatrix}, \mbox{\qquad}\phi^\dag(\vec{k}) = \begin{pmatrix} \vec{q}(-\vec{k}) && \vec{p}(-\vec{k}) \end{pmatrix}
\end{equation}

The equations of motion of $\phi(\vec{k})$ are given through the commutation relations, written here in compact matrix notation:
\begin{equation}
    \left[ \phi(\vec{k}), \phi^\dag(\vec{k}') \right] = - \Sigma_y \delta^3(\vec{k} -\delta{k'})
\end{equation}
where $\Sigma_y = \begin{pmatrix} 0 &&- i I_8 \\ i I_8 && 0 \end{pmatrix}$ encodes the canonical commutation relations of $q$ and $p$ and $I_8$ is the eight dimensional identity matrix. The equations of motion for $\phi(\vec{k})$ are then given by computing Heisenberg equations of motion:
\begin{equation}
    i \frac{d}{dt}\phi(\vec{k}) =  M(\vec{k}) \phi(\vec{k})
\end{equation}
where $M(\vec{k}) = - \Sigma_y A(\vec{k})$ is the dynamical matrix defined in equation (\ref{eq:eigenEQ}) of the main text. Looking for oscillating solutions in time $\phi(\vec{k},t) = \phi(\vec{k}) e^{-i \omega t}$ leads to the eigenvalue equation:
\begin{equation}
    \omega \vec{v}( \vec{k}, \omega) = M(\vec{k}) \vec{v}( \vec{k},\omega) 
\end{equation}
It can been proven\cite{xiao09theory} that for real eigenvalues the eigenvectors of $M(\vec{q})$ form an orthonormal basis under the inner product $-\Sigma_y$ i.e.
\begin{equation}
    \braket{v(\omega_i)}{v(\omega_j} \equiv  v(\omega_i)^\dag \cdot\left( - \Sigma_y \right)\cdot v(\omega_j) = |v(\omega_i)|^2 \delta_{i,j}
\end{equation}
Moreover, if $v(\omega)$ is an eigenvector with eigenvalue $\omega$ then $v(\omega)^*$ is also an eigenvector with eigenvalue $-\omega$ and opposite norm:
\begin{equation}
    v(\omega)^\dag \cdot\left( - \Sigma_y \right)\cdot v(\omega) = -  \left(v(-\omega)^\dag \cdot \left(- \Sigma_y\right) \cdot v(-\omega) \right)^*
\end{equation}
As a result one can order eigenvectors in such a way that positive frequency positive norm eigenvectors appear first and negative frequency negative norm eigenvectors appear second:
\begin{equation}
    T_d = \{ v(\omega_1), ..., v(\omega_n) , v(-\omega_1) ,  ..., v( -\omega_n) \}
\end{equation}
$T_d$ is the transformation matrix that relates canonically conjugated pairs to creation and annihilation operators:
\begin{equation}
    \phi(\vec{k}) = T_d(\vec{k}) \psi(\vec{k}),
\end{equation}
\newline
$\psi(\vec{k})$ contains the bosonic creation/annihilation operators that corresponds to each eigenvector:

\begin{equation}
\begin{split}
    \psi(\vec{k}) = \begin{pmatrix} b_1(\vec{k} ) \\
    \vdots \\
    b_1^\dag(- \vec{k}) \\
    \vdots
    \end{pmatrix} 
\end{split}
\end{equation}

The orthonormality condition on the eigenvectors can be summarized using the matrix equation
\begin{equation}
    T_d^\dag\cdot \left( - \Sigma_y\right) \cdot T_d = \Sigma_z
\end{equation}
with $\Sigma_z = \begin{pmatrix} I_4 && 0 \\ 0 && - I_4 \end{pmatrix}$, where we remind the reader that we include only eigenvectors that correspond to physical degrees of freedom which reduces the dimensionality of the system from 16 to 8. One can confirm that the above normalization conditions on the eigenvectors imply that $\psi(\vec{k})$ obeys the usual commutation relations:
\begin{equation}
    \left[ \psi(\vec{k}) , \psi^\dag(\vec{k}) \right] = \Sigma_z
\end{equation}
This mathematical structure underlies the majority of our work and allows us to project dynamics lowest energy subspace corresponding to JPs. Given an arbitrary function involving the state vector $\phi(\vec{k})$, dynamics can be projected to the JP subspace, $\{b_1, b_2, b_1^\dag , b_2^\dag\}$, by projecting each state vector inside the function:

\begin{subequations}
\begin{align}
    \mathcal{P}_{\rm JP} \cdot \phi(\vec{k}) =& c_1 b_1 + c_2 b_2 + c_1^\dag b_1^\dag + c_2^\dag b_2^\dag, \\
    c_i =& v(\omega_i(\vec{k})) \cdot \left( - \Sigma_y \right) \cdot \phi(\vec{k})
\end{align}
\end{subequations}

\section{Dispersion of plasmons in the non-relativistic limit}
\label{app:3plasmons}

\begin{figure}
    \centering
    \includegraphics[trim = {0cm 0cm 0cm 0cm},clip,scale =0.5]{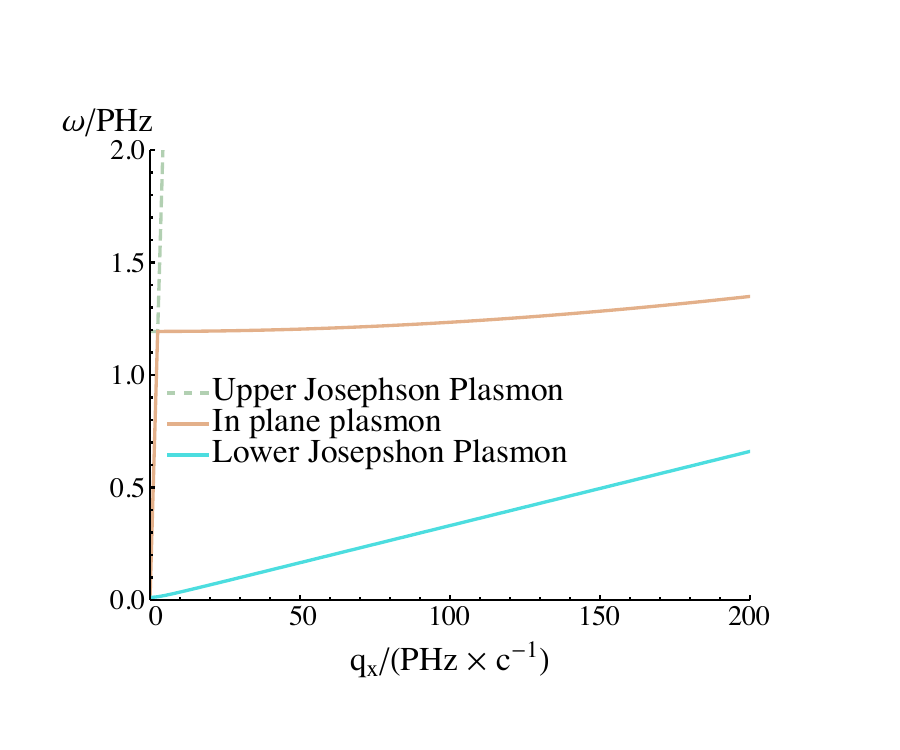}
    \caption{Dispersion of the three plasmons in the non-relativistic regime. The upper plasmon increases almost vertically due to the large speed of light. The remaining two plasmons increase linearly in energy as a function of $q_x$ for large $q_x$ due to the in plane kinetic energy contribution.}
    \label{fig:fulldisp}
\end{figure}

In this section, we show the dispersion of the system in the non-relativistic limit for all three plasmons in our system, shown in Fig.~\ref{fig:fulldisp}. Most notably, adopting a non-relativistic approach which replaces electric potentials with Coulomb's law ignores the upper JP along the $q_x$-axis which corresponds to a photon. Photon excitations are ignored in non-relativistic treatments because at large momenta $q_x >> \frac{1}{c} \sqrt{\Lambda_s \frac{\left(e^* \right)^2}{\epsilon}}$ the dispersion grows linearly with the large speed of light and these modes correspond very high energy excitations (seen in Fig.~\ref{fig:fulldisp} as a linear dispersion excitation with a large gradient in green). Our formalism interpolates between the relativistic regime presented in the main paper, which participates in the parametric amplification process and the non-relativistic regime which is often employed in solid state physics. In the non-relativistic limit, corrections from the in-plane kinetic energy offer significant contributions to the lower plasmon, which instead of saturating as predicted by previous work, it instead grows linearly.

\section{Four wave mixing parametric resonance}
\label{app:flucres}

In this Appendix we perform an instability analysis for four wave mixing process where one phonon acting as the external drive is converted into three JPs. This process is represented as a cubic interaction with a time-dependent coupling. We show in this Appendix that the same intuition can be used for four wave mixing as with three wave mixing processes. Instead of considering only equations of motion on one-point functions,e.g. $\expe{b}$, as is the case for normal parametric resonance, to treat a fluctuating interacting we need include two point functions as well, e.g. $\expe{bb}$, and perform a stability analysis around the equilibrium state. This amounts to performing an instability analysis around coherent state perturbations for which $\expe{b} \neq 0$, coupled to squeezed states for which $\expe{bb} \neq 0$. W.l.o.g. we consider a three mode Hamiltonian with non-linear driving that couples all three of them:
\begin{equation}
H = \sum_{i} \omega_i b^\dag_i b_i + 2 \epsilon \cos(\Omega t) \left( b^\dag_1 b^\dag_2 b^\dag_3+ h.c. \right).
\end{equation}
Notice that a particular form of cubic driving has been used including only terms that include three creation or three annihilation operators. One can confirm that all possible other combinations will not lead to an instability of the ground state. Linearizing around the undriven ground state, the equations of motion take the form:
\begin{subequations}
\begin{align}
\frac{d \expe{b_1^\dag }}{dt} =& i \omega_1 \expe{b^\dag_1} + i \epsilon 2 \cos\left( \Omega t \right) \expe{b_2 b_3}, \\
\begin{split}
\frac{d \expe{b_2 b_3} }{dt} =& - i \left( \omega_2 + \omega_3 \right)\expe{b_2 b_3} - \\ 
&i \epsilon 2 \cos\left( \Omega t \right) \expe{b_1^\dag} .
\end{split}
\end{align}
\label{eq:flucEOM}
\end{subequations}
These equations along with their permuted counter parts ($\{1,2,3\} \rightarrow \{3,1,2\} \rightarrow \{2,3,1\}$), will form the degenerate subspace in Floquet perturbation theory on resonance (see Section~\ref{sec:instability}):
\begin{equation}
\Omega = \omega_1 + \omega_2 + \omega_3
\end{equation}
Using the Floquet ansatz introduced in equation~(\ref{eq:floquetAnsatz}), the resulting secular equation has eigenfrequency
\begin{equation}
\omega = - \omega_2 \pm i \epsilon
\end{equation} 
As for three-wave mixing process, the time-dependent coupling leads to an instability on resonance with a growth rate that corresponds to the non-linear amplitude, $\epsilon$. For extended states, the non-linear amplitude is proportional to $\epsilon \propto \frac{1}{\sqrt{V}}$. This suppression makes it a subleading process compared to three-wave mixing terms.

Finally, we would like to stress that this process has no classical analogue. The resonant pair corresponds to a coherent state coupled to a squeezed state. However, since this occurs for every mode in the resonant trio independently and at the same rate, we don't expect a true squeezed state to emerge unless a specific coherent state has been biased by an experimental probe.

\section{Dressed state  }
\label{app:dressed}
In the main text, we report that, away form resonance, states are shifted in energy as a result of the drive according to a simple level attraction rule. The level attraction rule is stated as follows:
For two modes at frequencies $\omega_1(q)$ and $\omega_2(-q + Q)$ coupled through a drive with frequency $\Omega_{\rm fl}$ and momentum $Q$, the energies of the modes become UV-shifted when the condition $\omega_1(q) + \omega_2(-q + Q) < \Omega_{\rm fl}$ is satisfied and IR-shifted for states that obey $\omega_1(q) + \omega_2(-q + Q) < \Omega_{\rm fl}$. This a consequence of Floquet perturbation theory which can be applied away from resonance. Under parametric driving the Hamiltonian for the two coupled modes takes the form:
\begin{equation}
\begin{split}
    H_q =&  \omega_{1,q} b_{1,q}^\dag b_{1,q} + \omega_{2,q-Q} b_{2,Q-q}^\dag b_{2,Q-q} \\
    &+ \left( \lambda e^{ -i \Omega_{\rm fl} t}b^\dag_{q,1} b^\dag_{2,Q-q} + h.c. \right) 
\end{split}
\end{equation}
This Hamiltonian is turned into a time-independent Hamiltonian through a suitable rotating frame transformation:
\begin{equation}
    U = Exp\{ - i \Omega_{\rm fl} b^\dag_{1,q} b_{1,q} \}
\end{equation}
The effect of this transformation is given by
\begin{equation}
    U b_{1,q} U^\dag = e^{- i\Omega_{\rm fl} t} b_{1,q}
\end{equation}
and the rotating frame Hamiltonian is found to be:
\begin{equation}
\begin{split}
    H' =& U H U^\dag + i U \partial_t U^\dag, \\
    =& (\omega_{1,q} - \Omega_{\rm fl}) b_{1,q}^\dag b_{1,q} +( \omega_{2,q-Q} ) b_{2,Q-q}^\dag b_{2,Q-q} \\
    &+ \left( \lambda b^\dag_{q,1} b^\dag_{2,Q-q} + h.c. \right) 
\end{split}
\end{equation}
We now compute the dressing of the vacuum and excited modes due to the parametric drive. The vacuum is dressed as 
\begin{equation}
    \ket{\tilde{0}} = \ket{0} + \frac{g}{\Omega_{\rm fl} - \omega_{1,q} - \omega_{2,q-Q} } b^\dag_{1,q} b^\dag_{2,Q-q}\ket{0}
\end{equation}
The dressed vacuum energy becomes
\begin{equation}
    \widetilde{E}_0 = \frac{|g|^2}{\Omega_{\rm fl} - \omega_{1,q} - \omega_{2,q-Q} }
\end{equation}
While the first excited state for each mode becomes:
\begin{subequations}
\begin{align}
    \ket{\widetilde{1}}_{1,q} =& b_{1,q}^\dag \ket{0} + \frac{\sqrt{2} g}{\Omega_{\rm fl} - \omega_{1,q} - \omega_{2,q-Q} } \left( b_{1,q}^{\dag}\right)^2 b_{2,Q-q}^\dag \ket{0} \\
     \ket{\widetilde{1}}_{2,Q-q} =& b_{2,Q-q}^\dag \ket{0} + \frac{\sqrt{2} g}{\Omega_{\rm fl} - \omega_{1,q} - \omega_{2,q-Q} } \left( b_{2,Q-q}^{\dag}\right)^2 b_{1,q}^\dag \ket{0}
\end{align}
\end{subequations}
whose energy is shifted by:
\begin{subequations}
\begin{align}
    \widetilde{E}_{1,q}^{(1)} =& \omega_{1,q} + \frac{2 |g|^2}{\Omega_{\rm fl} - \omega_{1,q} - \omega_{2,-q}} , \\
    \widetilde{E}_{2,-q}^{(1)} =& \omega_{2,-q} + \frac{2 |g|^2}{\Omega_{\rm fl} - \omega_{1,q} - \omega_{2,-q}} 
\end{align}
\end{subequations}
As a result the renormalized energy of a single plasmon excitation in either band is shifted by the same amount,
\begin{equation}
    \widetilde{\omega}_i = \widetilde{E}^{(1)}_i- \widetilde{E}^{(0)} = \omega_i + \frac{|g|^2}{\Omega_{\rm fl} - \omega_{1,q} - \omega_{2,q-Q}}
\end{equation}
where $i = \{(1,q), (2,-q)\}$. Following a similar analysis it is easy to check that this applies to all higher energy states: 
\begin{equation}
    \widetilde{E}^{(n+1)}_i - \widetilde{E}^{(n)}_i = \omega_i + \frac{|g|^2}{\Omega_{\rm fl} - \omega_{1,q} - \omega_{2,-q}}
\end{equation}
From this analysis we conclude that generically due to a parametric drive, pair of states below parametric resonance, $\omega_{1,q} + \omega_{2,q-Q} < \Omega_{\rm fl} $ are UV-shifted while states above parametric resonance $\omega_{1,q} + \omega_{2,q-Q}> \Omega_{\rm fl}$ are IR-shifted.

\section{Floquet Fresnel problem}
\label{app:floqfresn}

In this section we will provide a general framework for calculating reflectivity in the presence of a periodic drive. We will keep the discussion general and allow for the periodic drive to have finite momentum. Reflectivity in a periodically driven system can be calculated by finding first the Floquet eigenstates oscillating at the frequency of the incident beam, $\omega_s$. Describing the Floquet drive by the effective Hamiltonian in equation~(\ref{eq:hamileffFloq}) is equivalent to starting from the equations of motion in the original variables with the a time periodic perturbation,
\begin{equation}
    i \partial_t \vec{v} = \left( M_0 + M_1 \cos\left( \Omega_{\rm fl} t \right)\right) \vec{v}
\label{eq:classicalFloq}
\end{equation}
and projecting the driving matrix, $M_1$, to the subspace of the JPs, which correspond to the lowest energy eigenstates of the equilibrium matrix $M_0$. Under a time-periodic drive, eigenvectors of the equations of motions have the Floquet form:
\begin{equation}
    \vec{v}(\vec{x}, t) = \sum_{i,n} e^{  i n \Omega_{\rm fl} t - i \omega t + i q_n x } v_{i} (\vec{q}_n)
\label{eq:floqExp}
\end{equation}
where $v_{i}(q_n)$ is a shorthand notation for the equilbrium eigenvectors that obey
\begin{equation}
    \omega_i(\vec{q}_n) \vec{v}_i (\vec{q}_n) = M_0 \vec{v}_{i} (\vec{q}_n), 
\end{equation}
and $i$ labels the plasmon branch. In equation~(\ref{eq:floqExp}), $\vec{q}_n$ reflects the fact that $M_1$ can in principle break translational invariance and mix states with different momenta, as is the case when the pump originates from a finite momentum JP instability. Assuming the driving amplitude is small, the above sum can be restricted to combination of modes that are coupled resonantly only within the nearest neighbouring Floquet bands. In other words, we include only the signal, $\omega_s$ and idler, $\omega_{\rm id} = \Omega_{\rm fl} - \omega_s$, frequency components in the ansatz to diagonalize equation~(\ref{eq:classicalFloq}),
\begin{widetext}
\begin{equation}
    \vec{v}_{i,\rm fl}(\vec{q}) = \alpha_s^i(\vec{q}) \vec{v}_i(\vec{q})e^{- i\omega_s t} + \alpha^i_{\rm idler}(\vec{q}) \left(\vec{v}_j\left(\vec{Q}-\vec{q}\right)\right)^*)e^{i \omega_{\rm id} t} 
\end{equation}
\end{widetext}
where $j$ and $\vec{Q}$ are determined by the matrix $M_1$. In particular, $\vec{Q}$ corresponds to the momentum of the pump. For effective Hamiltonians considered in the main text in equations~(\ref{eq:effLower})-(\ref{eq:effUpper}), the Floquet eigenstates involve a linear combination of two equilibrium states. The coefficients $\{\alpha^i_s(\vec{q}), \alpha_{\rm id}^i (\vec{q})\}$ obey the secular equation
\begin{equation}
    \begin{pmatrix} 
    - i( \omega_s - \omega_i(\vec{q}) ) && i g^* \\
    -i g && i (\omega_{\rm id} - \omega_j(
    \vec{q}-\vec{Q}) )
    \end{pmatrix}
    \cdot  
    \begin{pmatrix} \alpha_s^i (q) \\ \alpha^i_{\rm idler} (q)  \end{pmatrix} = 0 ,
\label{eq:Floqsec}
\end{equation}
where $g$ is proportional to the amplitude of the oscillating field and the matrix element of the overlap with matrix $M_1$. Solution of (\ref{eq:Floqsec}) is found by setting the determinant to zero,
\begin{equation}
  ( \omega_s - \omega_i(\vec{q}) )(\omega_{\rm id} - \omega_j(
    \vec{q}-\vec{Q}) ) = |g|^2
\label{eq:floquetDetapp}
\end{equation}
Equation~(\ref{eq:floquetDetapp}) provides an implicit equation for the wave-vector $\vec{q}$ of the transmitted wave when the incident beam oscillates at the signal frequency, $\omega_s$. In the normal incidence geometry consider in this paper (see Fig.~\ref{fig:reflStaticSchem}), $q_z = 0$. For small enough $|g|$, the above equation has two solutions, a signal solution $q_s$ for which $\omega_i(q_s) \approx \omega_s$ and an idler $q_{\rm id}$ for which $\omega_j( Q - q_{\rm id}) \approx \omega_{\rm id}$. The distinction is blurred close to parametric resonance when $\omega_i(q ) + \omega_j( Q - q) = \Omega_{\rm fl} $ and thus the two solutions will be denoted as $q = q_{\pm}$. From the secular equation, the Floquet eigenstate corresponding to the transmitted wave is given by:
\begin{equation}
\begin{split}
    v_{\rm fl, i}(q_\pm) =& E_0 t_{\pm}^i \bigg( g \vec{v}_i\left(\vec{q}\right) +\\
    & \left(\omega_s - \omega_i(\vec{q})\right)\left(\vec{v}_j\left(\vec{Q} - \vec{q}\right)\right)^*  \bigg)\\
    &
\end{split}
\end{equation}
where we set $\alpha^i_s(\vec{q}_\pm) = E_0 t^i_\pm g$, $\alpha^i_{\rm id}(\vec{q}_\pm) = E_0 t_{\pm}^i \left( \omega_s - \omega_{i}(\vec{q}_\pm) \right)$. The overall normalization of the Floquet eigenstate is absorbed in the transmission coefficients $\{t^{i}_{\pm} \}$ for each transmission channel and $E_0$ is the amplitude of the incident electric field. The explicit form of the electric potential in the transmitted wave is given by:
\begin{equation}
\begin{split}
&V_{\lambda,i}(\vec{x},t) = E_0 \sum_l t^l_\pm \big( \alpha_s^l(\vec{q}_{\pm}^{\mbox{ }l}) v^{V_{\lambda,i}}_{l}(\vec{q}_{\pm}^{\mbox{ }l}) e^{i q^l_{\pm} x} e^{-i \omega_s t} \\&+ \alpha_{\rm id}^{l}(\vec{q}_\pm^{\mbox{ }l})\left(v_{l'}^{V_{\lambda,i}}(\vec{Q} - \vec{q}_{\pm}^{\mbox{ }l})\right)^* e^{i\left( q^l_{\pm}- Q_x\right) x}e^{-i Q_z D i} e^{i \omega_{\rm id } t}\big)
\end{split}
\label{eq:transmexpApp}
\end{equation}
and similarly for $\vec{A}_{\vec{x}}$ and $A_z$. In equation~(\ref{eq:transmexpApp}), different independent transmission channels are included explicitly denoted by $l$, where transmission coefficients and wave-vectors depend on the particular transmission channel (see discussion of the static case in subsection~\ref{sec:statrefl}).

As in the static case the transmission electric and magnetic fields are calculated through equation~(\ref{eq:transStatic}), repeated here for convenience,
\begin{subequations}
\begin{align*}
    E_{z,\lambda,i} =& - \Delta_z V_{\lambda,i} - \partial_t A_{z,\lambda,i} ,\\
    B_{z,\lambda,i} =& - \partial_x A_{z,\lambda,i} + \Delta_z A_{x,\lambda,i}.
\end{align*}
\end{subequations}
Since the transmitted waves have components that oscillate at the idler frequency due to the Floquet drive, the reflected wave necessarily has components that also oscillate at the idler frequency to match boundary conditions. The out of plane momentum of the drive along the $c$-axis is conserved across the interface. The idler boundary conditions are given by
\begin{widetext}
\centering
\begin{subequations}
\begin{align}
    r_{\omega_{\rm idler}, Q_z + \frac{2 \pi n }{D}} E_0 =& E_{t,\omega_{\rm idler}, Q_z + \frac{2 \pi n }{D}}, \\
    -\frac{i \sqrt{ ( Q_z + \frac{2 \pi n }{D})^2 - \frac{\omega_{\rm idler}^2}{c_{vac}^2}   }}{\omega_{\rm idler}}r_{\omega_{\rm idler}, Q_z + \frac{2 \pi n }{D}} E_0 =& B_{t, \omega_{\rm idler}, Q_z + \frac{2 \pi n }{D}}
\end{align}
\end{subequations}
\label{eq:Floqbound}
\end{widetext}
Where we included the possibility for Bragg reflection as in equilibrium (discussed in subsection~\ref{sec:statrefl}). Schematically the reflected and transmitted waves in a Floquet medium are shown in Fig.~\ref{fig:floquetenhancement}. 

Finally, we note that one could simplify the system of equations even further by including only the transmitted channel corresponding to modes with the most overlap with the incoming beam (lower JP for signal frequencies $\omega_s < 10THz$ and upper JP for signal frequencies $\omega_s > 12 THz$). Consequently, to keep the number of variables the same as the boundary conditions, one needs to assume that Bragg reflection of the light is negligible and consider only the $n = 0$ case in equation~(\ref{eq:Floqbound}) and (\ref{eq:boundstat}). This approximation is valid when one of the transmission channels is either heavily localized at the interface as an evanescent wave within the frequency range considered or has a large in-plane current density component and hence small overlap with the electric field.

\subsection{Dissipation and causality}
Within our formalism, phenomenological dissipation is added through the dispersion relation by replacing $\omega_s \rightarrow \omega_s + i \gamma_i(q) $ where dissipation can depend on the particular transmission channel. Note that the above recipe implies for the idler frequency, $\omega_{id} \rightarrow \omega_{id} - i \gamma_i(q)$. Even in the absence of dissipation, an infinitesimal decay constant is needed to unambiguously determine the direction of the transmitted wave. The wave-vector solutions to the Floquet-Fresnel problem $q_\pm$ are chosen such that the (possibly infinitesimal) imaginary part is positive and therefore decays away from the surface and into the material. This procedure shows that the solution corresponding to the idler mode propagates towards the surface. This underlies parametric amplification, where energy can be extracted from the oscillating field in the form of waves transmitted outside of the material. The above situation is depicted schematically in Fig.~\ref{fig:floquetenhancement}.

\bibliographystyle{plain}
\bibliographystyle{unsrt}
\bibliography{references}{}

\begin{thebibliography}{10}
\bibitem{currentdeff20}
Note that we use notations for the inter-layer currents which are convenient in
  the discussion of lattice QED but differ from the usual notations in
  superconductivity. The inter-layer current in our notation corresponds to the
  physical current times the appropriate layer separation, which explains the
  extra factors of $1/d_\lambda$ in equation (\ref{J_zgrad_def} ) and factors
  $d_\lambda^2$ in equation (\ref{eq:curr}).
\bibitem{offresonant20}
Intuitively off-resonant contributions average to zero. Within perturbation
  theory off-resonant corrections are divided by their detuning making such
  contributions subleading. \cite{Rodriguez18}.
\end{thebibliography}

\end{document}